\documentclass[11pt,preprint]{aastex}

\shorttitle{Lack of dark matter it the solar neighborhood}
\shortauthors{Moni Bidin et~al.}

\begin{document}

\title{Kinematical and chemical vertical structure of the Galactic thick disk\thanks{Based on observations collected at the
European Organisation for Astronomical Research in the Southern Hemisphere, Chile
(proposal IDs 075.B-0459(A),077.B-0348(A))}~$^{,}$\thanks{This paper includes data gathered with the 6.5-meter Magellan and the
duPont Telescopes, located at Las Campanas Observatory, Chile.} \\
II. A lack of dark matter in the solar neighborhood}

\author{C. Moni Bidin}
\affil{Departamento de Astronom\'ia, Universidad de Concepci\'on, Casilla 160-C, Concepci\'on, Chile}
\email{cmbidin@astro-udec.cl}
\author{G. Carraro\altaffilmark{1}}
\affil{European Southern Observatory, Alonso de Cordova 3107, Vitacura, Santiago, Chile}
\altaffiltext{1}{Dipartimento di Astronomia, Universit\'a di Padova, Vicolo Osservatorio 3, I-35122, Padova, Italia}
\author{R. A. M\'endez}
\affil{Departamento de Astronom\'ia, Universidad de Chile, Casilla 36-D, Santiago, Chile}
\and
\author{R. Smith}
\affil{Departamento de Astronom\'ia, Universidad de Concepci\'on, Casilla 160-C, Concepci\'on, Chile}

\begin{abstract}
We estimated the dynamical surface mass density $\Sigma$ at the solar position between $Z$=1.5 and 4~kpc from the Galactic
plane, as inferred from the kinematics of thick disk stars. The formulation is exact within the limit of validity of a few
basic assumptions. The resulting trend of $\Sigma (Z)$ matches the expectations of visible mass alone, and no dark
component is required to account for the observations. We extrapolate a dark matter (DM) density in the solar neighborhood
of 0$\pm 1$~mM$_\odot$~pc$^{-3}$, and all the current models of a spherical DM halo are excluded at a confidence level
higher than 4$\sigma$. A detailed analysis reveals that a small amount of DM is allowed in the volume under study by the
change of some input parameter or hypothesis, but not enough to match the expectations of the models, except under an
exotic combination of non-standard assumptions. Identical results are obtained when repeating the calculation with
kinematical measurements available in the literature. We demonstrate that a DM halo would be detected by our method, and
therefore the results have no straightforward interpretation. Only the presence of a highly prolate (flattening $q>$2) DM
halo can be reconciled with the observations, but this is highly unlikely in $\Lambda$CDM models. The results challenge the
current understanding of the spatial distribution and nature of the Galactic DM. In particular, our results may indicate
that any direct DM detection experiment is doomed to fail, if the local density of the target particles is negligible.
\end{abstract}



\section{INTRODUCTION}
\label{s_intro}

Today, it is widely accepted that a dominant fraction of the mass in the universe is in the form of a
non-luminous (dark) matter (DM), whose nature is still unknown. Despite this general agreement, and decades of
investigations, its spatial distribution and its density in the solar neighborhood ($\rho_{\odot,DM}$) are still
poorly constrained by the observations. The local density\footnote{The DM mass will be given in mM$_\odot$ throughout
the paper, where 1~mM$_\odot =10^{-3}$M$_\odot =0.038$~GeV.} extrapolated from the Milky Way rotation curve
and other observational data, assuming a spherical DM halo, is in the range 5--13~mM$_\odot$~pc$^{-3}$
\citep{Weber10}. However, $\rho_{\odot,DM}$ could be significantly different in the case of an oblate or prolate halo,
or in the presence of a DM disk. Our poor knowledge of the DM density distribution is very unfortunate, since
this information is crucial to clarify the nature and properties of DM itself. The shape of the dark halo,
quantified by its shortest-to-longest axis ratio $q$, is in fact an important diagnostic on the nature of the
DM: for example, a round halo ($q\approx$1) is expected by hot DM models \citep{Peebles93}, while a very flat halo
($q\approx$0.2) is preferred if cold molecular gas or massive decaying neutrinos are the main constituent of the DM
\citep{Pfenniger94,Sciama90}. Noticeably, great effort is currently spent in experiments for direct DM
detection. The results of these experiments are degenerate between the unknown interaction cross-section of the
searched particles and their local density \citep[e.g.,][]{Gaitskell04,Aprile05}. Thus, most of the works estimating
the properties of the Weakly Interacting Massive Particles (WIMPs) have so far drawn their conclusions assuming the
local DM density of the Standard Halo Model \citep[SHM,][]{Jungman96}, $\rho_{\odot,DM}=8$~mM$_\odot$~pc$^{-3}$.
Clearly, observational constraints on the DM density at the solar position and on the flattening of the dark halo are
key components for revealing the secrets of the Galactic DM.

``Weighing'' the Galactic disk by means of the spatial distribution and kinematics of its stellar component is an
ancient art, dating back nearly one century \citep{Kapteyn22,Oort32}. The difference between the measured
mass and the visible mass provides an estimate of the amount of DM in the volume under analysis, and
constraints on the shape of the DM halo can be derived. The fundamental basis for this classical measurement is
the application of the Poisson-Boltzmann and Jeans equations to a virialized system in steady state. This allows us
to estimate either the local density at the solar position or the surface density (mass per unit area) of the mass
within a given volume. \citet{Garrido10} recently argued against the reliability of this method, but \citet{Sanchez11}
demonstrated that it can be applied to the Galactic disk, because it is in equilibrium with the Galactic potential. The
measurements of the dynamical mass in the solar neighborhood are abundant in the literature, and all but few works
\citep{Bahcall84,Bahcall92} came to the same overall conclusion that ``{\it there is no evidence for a significant
amount of disk DM}''. Different interpretations of this statement have been presented though: while the measurements
of the surface density usually match the expectations of the visible mass plus a classical spherical DM halo
\citep{Kuijken89a,Flynn94,Siebert03,Holmberg04,Bienayme06}, the estimates of the local volume density usually find a
much lower quantity of DM \citep{Kuijken89b,Creze98,Holmberg00,Korchagin03,deJong10}, although they are
compatible with the presence of a classical halo within the uncertainties. Despite many decades of investigation,
little progress has been made beyond this statement: for example, the most recent measurements are still compatible
with both the expectations of the SHM and the complete absence of DM in the solar neighborhood
\citep[e.g.,][]{Garbari11}. Some weak constraints on the properties of the DM halo were given by \citet{Creze98} and
\citet{Bienayme06}, who claimed that a DM halo flattening $q\leq 0.5$ is excluded by the observations.

The main limitations to the measurements of the dynamical mass are imposed by the great observational effort required,
because the information about the kinematics and spatial distribution of a vast number of stars is needed. Some
approximations were thus always introduced in the formulation, accurate only at low Galactic heights, whose
validity has since been questioned in the literature. For example, \citet{Siebert08} and \citet{Smith09} found
that at only $Z$=1~kpc from the Galactic plane the potential is not separable in the radial and vertical coordinates,
and neglecting the non-diagonal term of the dispersion matrix could introduce a bias. \citet{Garbari11} found that
this becomes an issue at only $Z$=0.5~kpc, and argued against the use of the approximate formulation of
\citet{Holmberg00,Holmberg04} at high $Z$. As a consequence, all previous investigations were limited to
$Z\leq 1.1$~kpc from the Galactic plane. The amount of DM expected in this volume is small compared to observational
uncertainties, and no strong conclusion could be drawn. Moreover, only \citet{Korchagin03} directly calculated the mass
density from an analytical expression, while the other investigations estimated it by comparison of the observational
quantities with the expectations of a Galactic mass model.

In this paper, we propose a new approach to estimate the surface mass density of the Galactic disk. Our
formulation leads to an analytical expression that is exact within the limits of validity of a few basic assumptions.
The surface density can thus be directly calculated at any distance from the Galactic plane. The calculation requires
a knowledge of the spatial variations of the three-dimensional kinematics of a test stellar population. Our calculations
are based on the results of our recent investigation of the Galactic thick disk kinematics \citep{Carraro05,Moni09}.
Similar results were presented by \citet{Moni10}, who searched for a signature of the dark matter disk, predicted by
the merging scenario of thick disk formation.


\section{THE THEORY}
\label{s_theory}

\renewcommand{\theenumi}{\Roman{enumi}}
\renewcommand{\labelenumi}{\theenumi}

In the following, we will use the cylindrical Galactic coordinates ($R,\theta,Z$), where $R$ is the Galactocentric
distance, $\theta$ is directed in the sense of Galactic rotation, and $Z$ is positive toward the north Galactic
pole. The respective velocity components are ($\dot{R},\dot{\theta},\dot{Z}$)=($U,V,W$).

Our formulation is based on the integrated Poisson equation in cylindrical coordinates
\begin{equation}
-4\pi G\Sigma(Z) = \int_{-Z}^{Z} \frac{1}{R}\frac{\partial}{\partial R}(R F_{R}) dz +2\cdot [F_{z}(Z) - F_{z}(0)],
\label{e_Pois}
\end{equation}
where G is the gravitational constant, $\Sigma (Z)$ is the surface density of the mass comprised between $\pm Z$
from the Galactic plane, and the radial and vertical component of the force per unit mass, F$_\mathrm{R}$ and
F$_\mathrm{Z}$ respectively, can be expressed through the Jeans equations
\begin{equation}
F_{R}=-\frac{\partial \phi}{\partial R}=\frac{1}{\rho}\frac{\partial (\rho\overline{U^2})}{\partial R}+
\frac{1}{\rho}\frac{\partial (\rho\overline{UW})}{\partial Z}+\frac{\overline{U^2}-\overline{V^2}}{R}+
\frac{1}{\rho}\frac{\partial (\rho \overline{U})}{\partial t},
\label{e_JeansR}
\end{equation}
\begin{equation}
F_{Z}=-\frac{\partial \phi}{\partial Z}=\frac{1}{\rho}\Bigl{[}\frac{\partial (\rho\overline{W^2})}{\partial Z}+
\frac{\rho\overline{UW}}{R}+\frac{\partial (\rho\overline{UW})}{\partial R}+
\frac{\partial (\rho\overline{W})}{\partial t}\Bigl{]},
\label{e_Jeansz}
\end{equation}
where $\rho$ is the volume density, and $\phi$ is the gravitational potential.
The following basic assumptions can be made, for symmetry reasons, for the stellar population under study:
\begin{enumerate}
\item The test population is a virialized system in steady state. All the temporal derivatives are therefore null:
\begin{equation}
\frac{\partial (\rho \overline{U})}{\partial t}=\frac{\partial (\rho\overline{W})}{\partial t}=0
\label{eq_steady}.
\end{equation}
\label{h_steady}
\item The vertical component of the force per unit mass is null on the plane:
\begin{equation}
F_{Z}(0)=0
\label{e_force}.
\end{equation}
\label{h_force}
\item The trend of velocity dispersions with $Z$ is symmetric with respect to the plane:
\begin{equation}
\sigma_i(Z)=\sigma_i(-Z)
\label{e_simm},
\end{equation}
with $i=U,V,W$.
\label{h_simm}
\item $\overline{UW}$ is antisymmetric with respect to the Galactic plane:
\begin{equation}
\overline{UW}(Z)=-\overline{UW}(-Z)
\label{e_uw}.
\end{equation}
\label{h_uw}
\item The net radial and vertical bulk motion of the trace population is null:
\begin{equation}
\overline{U}=\overline{W}=0
\label{e_net}.
\end{equation}
\label{h_net}
\end{enumerate}
\citet{Sanchez11} demonstrated the validity of assumption~(\ref{h_steady}), that was recently questioned in the
literature \citep{Garrido10}. The hypothesis~(\ref{h_force}), (\ref{h_simm}), and (\ref{h_uw}) are required for
symmetry reasons, and they imply that the integral of the velocity dispersions in $\pm Z$ is twice the integral
between zero and $Z$, while the integral of $\overline{UW}$ is null, and $\overline{UW}$(0)=0. It can be easily
seen that assumption~(\ref{h_uw}) directly results from the Jeans equations, if the symmetry requirements
F$_{R}(Z)$=F$_{R}(-Z)$ and F$_{Z}(Z)=-$F$_{Z}(-Z)$ are to be satisfied. \citet{Moni10} also showed that the
cross-term must be assumed anti-symmetric with respect to $Z$, else the calculation leads to unphysical results.
Recent investigations detected a non-null mean radial motion of stars \citep{Dinescu11,Moni12}. However, this is
not a consequence of a global motion of disk stars, but it is rather a signature of local kinematical
substructures, so that assumption~(\ref{h_net}) is still valid as a general property of disk stars.

Inserting Equations~(\ref{e_JeansR}) and (\ref{e_Jeansz}) into Equation~(\ref{e_Pois}), and making use of the
assumptions~(\ref{h_steady}) to (\ref{h_net}), the resulting equation can be analytically solved for $\Sigma(Z)$
at any $R$ and $Z$, if the three-dimensional analytical expressions for the kinematics and the mass density
distribution of the test population are given. To provide this extensive input, we make use of observational
results only. The spatial distribution of the test population is fixed by the following assumptions:
\begin{enumerate}
\setcounter{enumi}{5}
\item The volume density of the test population decays exponentially both in the radial and vertical
direction, with exponential scale height and length h$_{Z,\rho}$ and h$_{R,\rho}$, respectively:
\begin{equation}
\rho (R,z)=\rho_\odot\cdot\exp{\Bigl{(}-\frac{Z-Z_\odot}{h_{Z,\rho}}-\frac{R-R_\odot}{h_{R,\rho}}\Bigl{)}}
\label{e_expo},
\end{equation}
where $\rho_\odot$ is the density at ($R,Z)=(R_\odot,Z_\odot$).
\label{h_expo}
\item The scale length of the test population, h$_{R,\rho}$, is invariant with respect to Galactic height:
\begin{equation}
\frac{\partial h_{R,\rho}}{\partial Z}=0
\label{e_noflare}.
\end{equation}
\label{h_cabrera}
\end{enumerate}
These hypotheses are a classical representation of Galactic disk-like populations, and (\ref{h_cabrera}) is
confirmed by the empirical results of \citet{Cabrera07}. Theoretical considerations suggest that the vertical
density profile should be much closer to a sech$^{2}$ function than an exponential \citep{Camm50,Camm52}, but
observational evidences confirm that the Galactic thick disk, object of our investigation, is well described by
the double exponential law of Equation~(\ref{e_expo}) \citep[e.g.,][]{Hammersley99,Siegel02,Juric08}. The assumed
vertical decay is surely an accurate fit of the disk density at our large Galactic heights, where even the
sech$^{2}$ profile approximates to an exponential decline.

The calculation is simplified by two additional working hypothesis:
\begin{enumerate}
\setcounter{enumi}{7}
\item The rotation curve is locally flat in the volume under study:
\begin{equation}
\frac{\partial \overline{V}}{\partial R}=0
\label{e_flat}.
\end{equation}
\label{h_flat}
\item The disk test population is not flared in the volume of interest:
\begin{equation}
\frac{\partial h_{Z,\rho}}{\partial R}=0
\label{e_noflare}.
\end{equation}
\label{h_flare}
\end{enumerate}
These are very common assumptions, whose function in our approach is neglecting few small terms, which
complicate the formulation while introducing only second-order corrections. The observational evidences in
their support will be discussed in Section~\ref{s_an_noflat} and \ref{s_an_flare}, where the influence of their
break-down will also be analyzed.

It is easily demonstrated that, by means of our assumptions, the final expression for the surface density
reduces to
\begin{eqnarray}
\nonumber
-2\pi G\Sigma(R,Z) = \frac{\partial \sigma^2_W}{\partial Z}-\frac{\sigma^2_W}{h_{Z,\rho}} -
\int_{0}^{Z}\frac{\sigma^2_U}{Rh_{R,\rho}} dz + \int_{0}^{Z}\frac{\partial^2 \sigma^2_U}{\partial R^2} dz +
\int_{0}^{Z}\Bigl{(}\frac{2}{R}-\frac{1}{h_{R,\rho}}\Bigl{)}\cdot\frac{\partial \sigma^2_U}{\partial R} dz - \\
- \frac{1}{R}\int_{0}^{Z}\frac{\partial \sigma^2_V}{\partial R} dz +
\overline{UW}\Bigl{(}\frac{2}{R}-\frac{1}{h_{R,\rho}}\Bigl{)} + 2\frac{\partial \overline{UW}}{\partial R} -
\frac{1}{h_z\rho}\int_{0}^{Z}\frac{\partial \overline{UW}}{\partial R} dz.
\label{e_general}
\end{eqnarray}
This equation is exact within the limits of validity of the underlying assumptions, and it can be used to calculate
the surface density of the mass within $\pm Z$ from the Galactic plane at the Galactocentric distance $R$. The
estimate requires the knowledge of the two parameters h$_{Z,\rho}$ and h$_{R,\rho}$, the vertical trend of the three
dispersions and of $\overline{UW}$, and the radial derivative of $\sigma_\mathrm{U}(Z)$, $\sigma_\mathrm{V}(Z)$,
and $\overline{UW}(Z)$. A great observational effort, impossible until the last decade, is required to collect
this quantity of information. Nevertheless, this will be eventually gathered by modern extensive surveys
such as the Sloan Digital Sky Survey \citep[SDSS;][]{York00}, GAIA \citep{Wilkinson05}, and Large Synoptic Survey
Telescope \citep[LSST;][]{Ivezic08}. The measurement of the mass distribution in the Galaxy will be then possible
with unprecedented detail.

As shown in Section~\ref{s_res}, the data used in the present study provide no information about the radial behavior
of the kinematics, and we are forced to model it with an additional assumption. We adopt the most likely hypothesis:
\begin{enumerate}
\setcounter{enumi}{9}
\item $\overline{UW}$ and the square of the dispersions exponentially decay in the radial direction, with the
same scale length as that of the volume mass density.
\label{h_radial}
\end{enumerate}
The effects of alternative radial trends will be analyzed in Section~\ref{s_an_param}. Assumption~(\ref{h_radial})
is observationally confirmed only for $\sigma_\mathrm{W}$ \citep{Kruit81,Kruit82}, and
its extension to the other components relies on the assumption of radially constant anisotropy, i.e. the radial
constancy of the ratio of the dispersions. However, both theoretical calculations and observations support this
hypothesis: \citet{Cuddeford92} demonstrated, by means of numerical integration of orbits, that it is the best
representation of the radial trend of dispersions for $R\leq$9~kpc, and the most detailed observations available
to date also confirm it \citep{Lewis89}, although \citet{Neese88} prefer a linear radial decay for
$\sigma_\mathrm{U}$. \citet{Moni12} also find that the observational data are consistent with
assumption~(\ref{h_radial}).

Using assumption~(\ref{h_radial}), Equation~\ref{e_general} can be simplified substantially, and after a
calculation involving only simple integrals and derivatives, we obtain
\begin{equation}
\Sigma(Z)=\frac{1}{2\pi G}\Bigl{[}
k_1\cdot\int_{0}^{Z}\sigma_{U}^{2}dz+k_2\cdot\int_{0}^{Z}\sigma_{V}^{2}dz+k_3\cdot \overline{UW}
+\frac{\sigma^{2}_{W}}{h_{Z,\rho}}-\frac{\partial \sigma^{2}_{W}}{\partial Z}\Bigl{]}
\label{e_final},
\end{equation}
where
\begin{eqnarray}
&&k_1=\frac{3}{R_\odot\cdot h_{R,\rho}}-\frac{2}{h_{R,\rho}^{2}}, \\
&&k_2=-\frac{1}{R_\odot\cdot h_{R,\rho}}, \\
&&k_3=\frac{3}{h_{R,\rho}}-\frac{2}{R_\odot}.
\end{eqnarray}


\section{RESULTS}
\label{s_res}

\begin{figure}
\epsscale{0.65}
\plotone{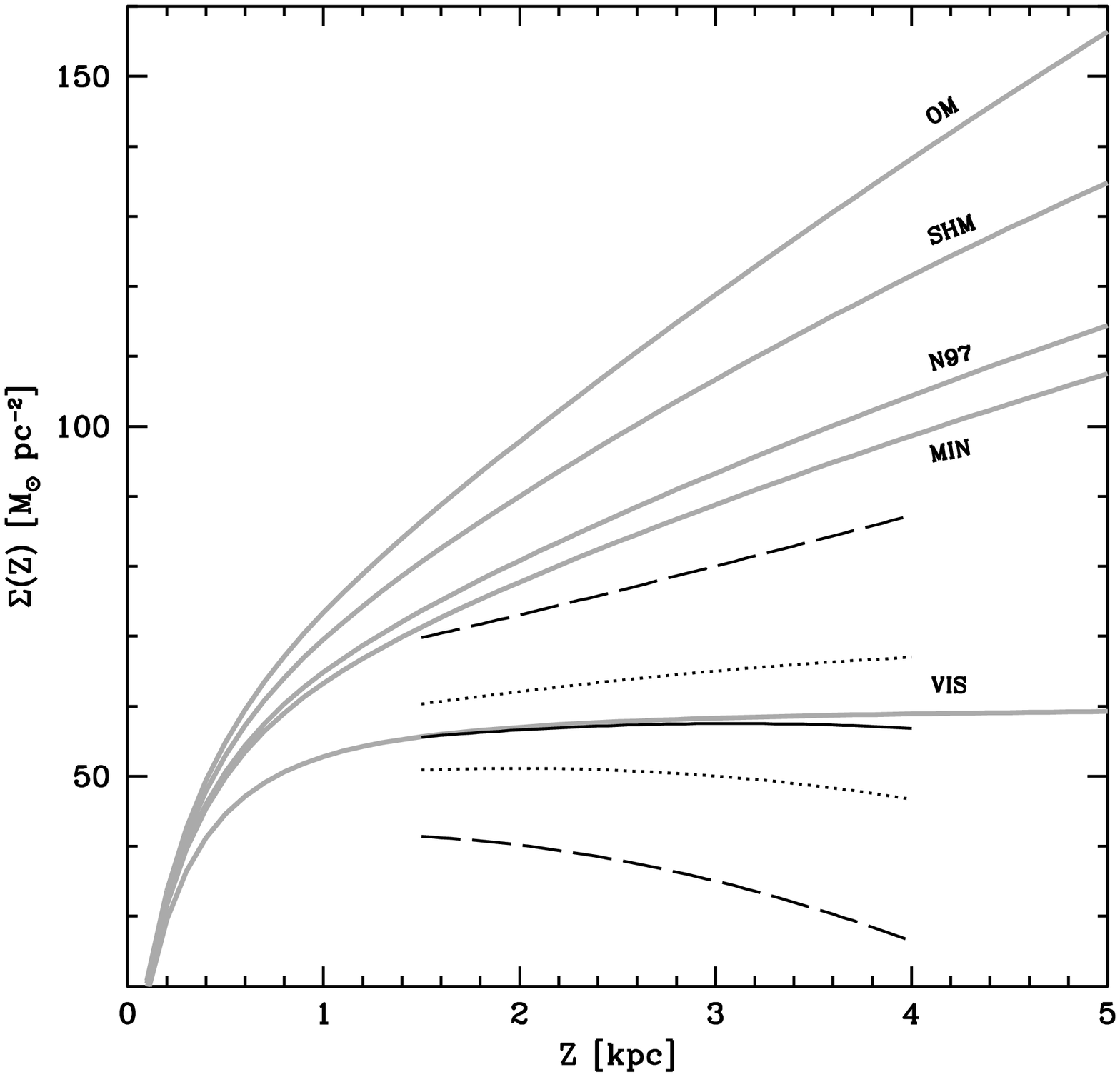}
\plotone{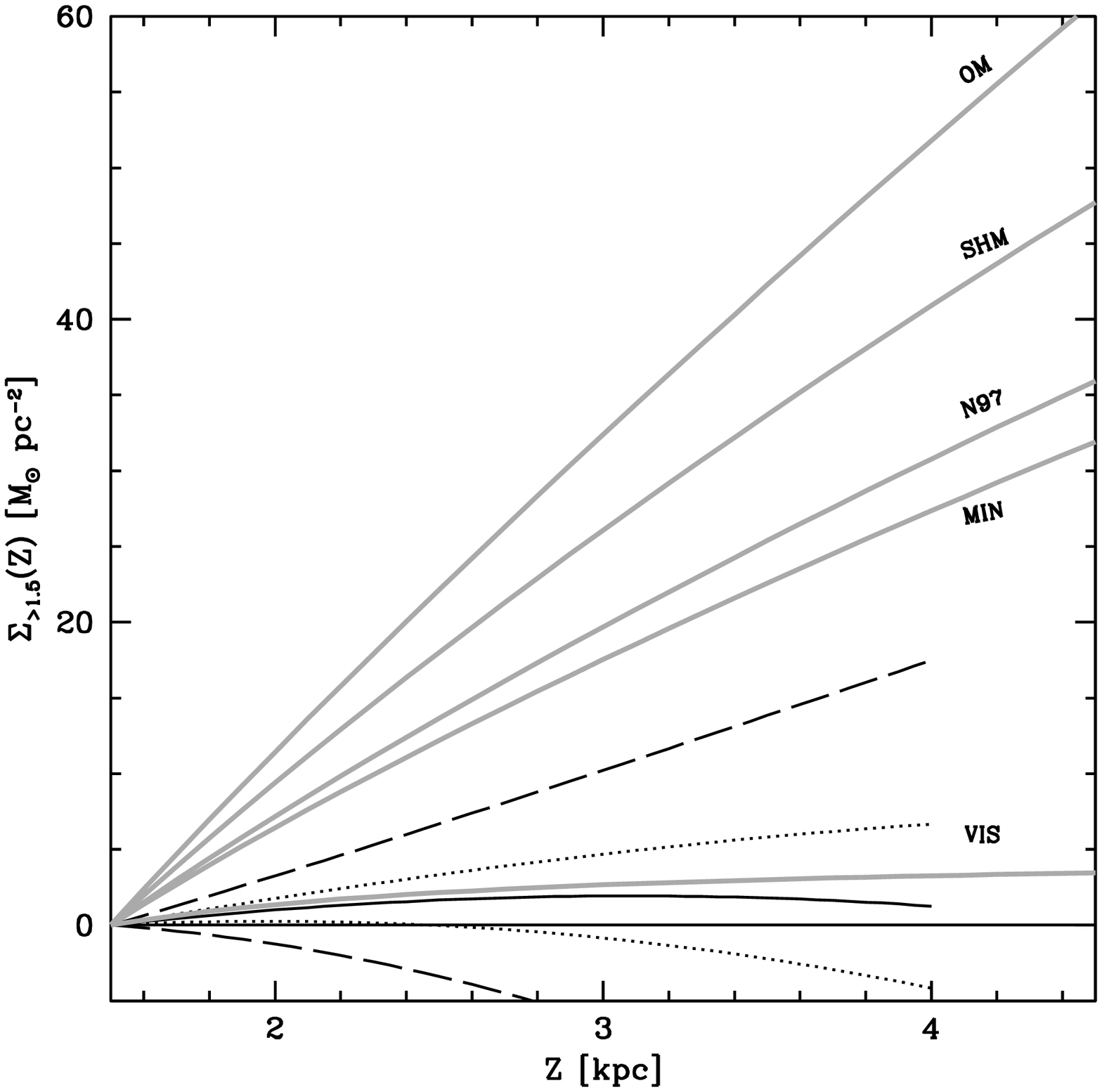}
\caption{Observational results for the absolute (upper panel) and incremental (lower panel) surface mass density,
as a function of distance from the Galactic plane (black curves), compared to the expectations of the models
discussed in the text (thick grey curves). The dotted and dashed lines indicate the observational 1$\sigma$ and
3$\sigma$ strip, respectively.
\label{f_resgen}}
\end{figure}

Our estimate of the surface mass density is based on the results of \citet{Moni12}, who measured the kinematics
of the Galactic thick disk, and its variation with distance from the plane, between $Z$=1.5 and 4.5~kpc. In brief,
they studied a sample of $\sim$400 thick disk red giants toward the south Galactic pole, vertically distributed
with respect to the Galactic plane, and derived their three-dimensional kinematics by means of 2MASS photometry
\citep{Skrutskie06}, SPM3 absolute proper motions \citep{Girard04}, and radial velocities \citep{Moni09}. Thus,
the data do not provide any information about the variation of the kinematics with Galactocentric distance, that
will be modeled with assumption~(\ref{h_radial}). The surface mass density will therefore be calculated by means
of Equation~(\ref{e_final}). \citet{Moni12} detected a clear increment with $Z$ of all the dispersions, well
represented by a linear fit. We will adopt for $\sigma_\mathrm{U}(Z)$, $\sigma_\mathrm{V}(Z)$, and
$\sigma_\mathrm{W}(Z)$ the relations given by equations~(3)--(5) of \citet{Moni12}. The vertical trend of
$\overline{UW}$ will be taken from the linear fit shown in Figure~2 of \citet{Moni10}, which yields the solution
\begin{equation}
\overline{UW}=(1522\pm 100)+(366\pm 30)\cdot (Z-2.5)~~~\mathrm{km^2~s^{-2}},
\label{e_cross}
\end{equation}
where $Z$ is in kpc.

Measuring the kinematical properties and the spatial distribution of the test population by means of the
same observed stars would be highly desirable. Unfortunately, our sample is not suitable for this: even if
\citet{Girard06} used it to estimate the thick disk scale height, their measurement could have been affected by the
thin disk contamination unaccounted for in their study \citep{Moni12}. However, our sample can be considered
representative of the thick disk kinematics at the solar position, whose spatial distribution has been extensively
studied in the last three decades. This would not be true if the thick disk was a mixture of sub-populations with
different properties. In this case, the spatial distribution of the sub-population under study, or the dominant one,
should be preferred. Nevetheless, to our knowledge, there is no evidence in literature supporting such high degree of
inhomogeneity in the thick disk, nor that the sample under study is for some reason peculiar. Finally, we fix
$R=R_\odot=8.0\pm$0.3~kpc, and h$_{R,\rho}=3.8\pm$0.2~kpc, h$_{Z,\rho}=900\pm$80~pc from the average of sixteen
and twenty-one literature measurements respectively\footnote{\citet{Ratnatunga89,Yamagata92,Hippel93,Beers95,Larsen96,
Robin96,Spagna96,Ng97,Buser99,Ojha99,Chiba00,Chen01,Lopez02,Siegel02,Larsen03,Cabrera05,Girard06,Vallenari06,Brown07,
Cabrera07,Arnatoddir08,Bilir08,Juric08,Veltz08,deJong10,Just11,Chang11}}. The impact of these parameters on the final
results will be analyzed in Section~\ref{s_an_param}. The errors were defined, as in \citet{Moni10}, from the
statistical error-on-the-mean. If the literature estimates can be considered independent measurements of an underlying
quantity, this must be the most rigorous estimate of the true uncertainty. However, the scale height and length of the
Galactic thick disk are traditionally poorly constrained, although the measurements converged considerably in the
last years. For this reason, \citet{Moni10} also considered enhanced errors (0.4 and 0.12~kpc for h$_{R,\rho}$ and
h$_{Z,\rho}$, respectively), possibly more representative of the true uncertainties. In this case, the final errors
on $\Sigma (Z)$ are enhanced by about 50\%, and the significance of the results thus decreases by a factor of
$\sim$0.7. Nevertheless, adopting these larger errors would not affect the general conclusions of our work noticeably
because, as will appear clearer later, the significance of the results remains very high even in this case.

The error on $\Sigma (Z)$ was calculated from the propagation of the uncertainties on the four kinematical
quantities and the three parameters. The resulting equation is particularly cumbersome, but it can be obtained
through simple derivations of the terms on the right hand side of Equation~(\ref{e_final}). Monte-Carlo simulations,
kindly provided us by the referee, confirmed that the final error is a good estimate of the uncertainty propagated
from the ones associated to the quantities entering in the calculations. The simulations were performed repeating
the estimate of $\Sigma (Z)$ after varying the input quantities, each one randomly drawn from a Gaussian
distribution with mean and dispersion given by the assumed value and error, respectively.

The resulting vertical profile of the surface mass density $\Sigma (Z)$, is shown in the upper panel of
Figure~\ref{f_resgen}. We find $\Sigma$(1.5~kpc)=55.6$\pm$4.7~M$_\odot$~pc$^{-2}$, and the profile is nearly flat,
increasing by only $\sim$2~M$_\odot$~pc$^{-2}$ up to $Z$=3~kpc and eventually bending down, while the error
monotonically increases up to 10~M$_\odot$~pc$^{-2}$ at $Z$=4~kpc. The decrease of $\Sigma (Z)$ in the last kpc is
unphysical, but it is negligible compared to the errors, corresponding to only 0.07$\sigma$. This could be
corrected by a tiny change of the input parameters, for example increasing h$_{R,\rho}$ by 0.1~kpc, but we will
not manipulate them to avoid the introduction of arbitrariness in the results. If the solution is extrapolated
to $Z$=0, the unphysical result $\Sigma (0)\neq 0$ is obtained. This is due to the fact that the vertical trend of
the kinematical quantities was derived from a linear fit. This is a good approximation in the $Z$-range under
analysis, but the dispersions should depart from it at lower heights, bending down with steeper gradient. The most
clear case is $\overline{UW}$, which should be zero on the plane for symmetry reasons, but the linear fit returns
$\overline{UW}(0)=607$~km$^2$~s$^{-2}$. The extrapolation of the results thus fails to account for the changed
gradient of the kinematical quantities and, as a consequence, of $\Sigma (Z)$. In fact, as shown later
(Figure~\ref{f_din}), a steeper curve of the surface mass density at lower $Z$ (and a lower extrapolation to $Z$=0)
is recovered when using the steeper gradients of kinematics of \citet{Dinescu11}. It can be noted that the linear
fit is a worse approximation in this case, as reflected by $\Sigma (Z)$ being too flat compared to any model,
because the dispersions vary their gradient more quickly nearer to the plane. In conclusion, the extrapolation of
our results outside the $Z$-range where they were obtained is not allowed.

In Figure~\ref{f_resgen}, the results are compared with the expectations for the known visible mass, indicated by
the thick grey curve labeled as VIS. This was modeled as the sum of a thin layer of interstellar medium (ISM) of
13~M$_\odot$~pc$^{-2}$ \citep{Holmberg00}, plus three stellar components, the halo, the thin disk, and the thick
disk, whose geometrical parameters were taken from \citet{Juric08}. Their local density at $Z$=0 was normalized so
that the total surface density of the stellar disk is $\Sigma_\mathrm{disk}$(1.1~kpc)=40~M$_\odot$~pc$^{-2}$
\citep{Holmberg04,Bienayme06}. This is currently the best estimate assumed in Galactic mass models
\citep[e.g.,][]{Dehnen98,Olling01,Weber10}, but the local mass density of both the ISM and the stellar component
are still affected by observational uncertainties. However, as discussed below, the exact value of these parameters
does not have a significant influence on our results.

The estimate of the surface mass density matches the expectation of visible mass alone, and the degree of overlap
between the two curves is striking. There is no need for any dark component to account for the results: the measured
$\Sigma (Z)$ implies a local DM density $\rho_{\odot,DM}=0\pm1$~mM$_\odot$~pc$^{-3}$.  This estimate negligibly
changes ($\rho_{\odot,DM}=0.4\pm1.2$~mM$_\odot$~pc$^{-3}$) if we assume
$\Sigma_\mathrm{disk}$(1.1~kpc)=35~M$_\odot$~pc$^{-2}$ in the visible mass model \citep{Holmberg00,Garbari11}.
Fitting a DM halo model to the observations is pointless because, disregarding its exact shape, the procedure
necessarily converges to a zero-density solution. The more complex task of building a model able to reproduce our
results and other observational constraints (e.g. the Galactic rotation curve, the gas disk flare) is beyond the
scope of this paper. Nevertheless, we can compare the measurements with the expectations of the most popular DM halo
models.

A great quantity of models have been proposed in the literature to describe the spatial distribution of the Galactic
DM, but only the mass between 0 and 4~kpc from the plane at $R=R_\odot$ is involved in the comparison with the
observations. Hence, the local density $\rho_{\odot,DM}$ and the halo flattening $q$ are the only relevant
parameters, while the exact functional form is not critical. Indeed, it is easy to see that different models with
the same $\rho_{\odot,DM}$ and $q$ are indistinguishable in terms of the expected quantity of DM in the volume under
analysis. We will therefore compare the observations with a set of selected models, chosen on the basis of the
expected $\rho_{\odot,DM}$. We will assume spherical models here ($q$=1), while the effects of varying the
flattening parameter will be analyzed in more detail in Section~\ref{s_an_flattening}.

\citet{Olling01} presented a family of self-consistent models, based on a classical non-singular isothermal spheroid:
\begin{equation}
\rho_\mathrm{DM}(R,Z)=\rho_c\Bigl{[}\frac{R^2_c}{R^2_c+R^2_\odot+(Z/q)^2}\Bigl{]},
\label{eq_DMOll}
\end{equation}
\citep[e.g.,][]{vanAlbada86,Kent87}, where the core radius $R_c$ and the central density $\rho_c$ depend on $q$,
so that the resulting rotation curve is independent of it \citep{Olling95}. As a conservative choice, we will
assume, among the solutions proposed by \citet{Olling01} with $R_\odot$=7.8--8.5~kpc and
$\Sigma_\mathrm{disk}$(1.1~kpc)$\geq$30~M$_\odot$~pc$^{-2}$, the model with the minimum local density
($\rho_{\odot,DM}\approx$10~mM$_\odot$~pc$^{-3}$, hereafter model OM), which have $R_c$=8.01~kpc and
$\rho_c$=20.6~mM$_\odot$~pc$^{-3}$.

A more general expression of the DM halo shape is
\begin{equation}
\rho_\mathrm{DM}(R,Z)=\rho_{\odot,DM}\cdot\Bigl{(}\frac{\sqrt{R^2+(Z/q)^2}}{R_\odot}\Bigl{)}^{-\alpha}
\cdot\Bigl{[}\frac{1+(\frac{\sqrt{R^2+(Z/q)^2}}{R_c})^{\beta}}{1+(\frac{R_\odot}{R_c})^{\beta}}\Bigl{]}^{-\gamma},
\label{eq_DMgen}
\end{equation}
where the indices ($\alpha, \beta, \gamma$) characterize the radial fall-off of the density distribution. Various
sets of indices have been proposed in the literature, suggesting either ``cuspy''
\citep[e.g.,][]{Navarro97,Moore99,Binney01,Ludlow09} or ``cored'' profiles
\citep[e.g.,][]{deBoer05,Narayan05,Gentile07,Salucci07,Oh08}. The most representative models were analyzed by
\citet{Weber10}, who fixed the best-fit parameters on the basis of the most recent observational constraints.
Among these, we will consider the model which better fits the observations (in terms of the lowest $\chi^2$), the
\citet[][NFW]{Navarro97} profile ($\alpha= \beta =1, \gamma =2$) with $R_c$=10.8~kpc and
$\rho_{\odot,DM}$=8.4~mM$_\odot$~pc$^{-3}$. This value coincides with the Standard Halo Model density usually
assumed in direct DM detection experiments \citep{Jungman96}, and it will be referred to as SHM. We will also
consider the NFW profile with the lowest local density ($R_c$=20~kpc, $\rho_{\odot,DM}$=6.1~mM$_\odot$~pc$^{-3}$,
hereafter model N97), and the model with the minimum local DM density, a pseudo-isothermal profile
\citep[$\alpha=0, \beta =2, \gamma =1$,][hereafter model MIN]{deBoer05} with $R_c$=5~kpc and
$\rho_{\odot,DM}$=5.3~mM$_\odot$~pc$^{-3}$. This value is usually assumed as a lower limit for the local DM
density \citep{Garbari11,Weber10}.

In Figure~\ref{f_resgen}, the measured vertical profile of the surface density is compared to the expectations of
the selected DM halo models. The OM model is excluded at the 8$\sigma$ level, while the SHM at the 6$\sigma$ level.
Even the MIN model, with the minimum local density extrapolated from the Galactic rotation curve \citep{Garbari11},
is 4.1$\sigma$ more massive than the observed curve. If we assume
$\Sigma_\mathrm{disk}$(1.1~kpc)=35~M$_\odot$~pc$^{-2}$ in the visible mass model, the results have a lower but
still very significant departure: even the MIN model would still be excluded at the 3.6$\sigma$ level.

It could be argued that integrating the linear vertical profiles of the velocity dispersions in the range
$Z$=0--4~kpc implicitly assumes their extrapolation down to $Z$=0, while they were measured only for $Z\geq$1.5~kpc.
Indeed, the vertical trend of the dispersions is not expected to be strictly linear, but to be shallower at higher
distance from the plane \citep[see, for example, the discussion of][]{Dinescu11}. Nevertheless, assuming for
$Z\leq$1.5~kpc the steeper relations measured by \citet{Dinescu11}, the resulting surface density decreases by only
0.5~M$_\odot$~pc$^{-2}$. Arbitrarily increasing all the gradients by a factor of two for $Z$=0--1.5~kpc,
$\Sigma (Z)$ decreases by 5.5~M$_\odot$~pc$^{-2}$, enhancing the disagreement with the DM halo models. In
conclusion, the integration of the profiles in the entire range is likely a good approximation, and it cannot be
the cause of the mismatch between the measurements and the model expectations.

The increment of the surface density between 1.5 and 4~kpc ($\Sigma_{>1.5}(Z)$) can be measured integrating
Equations~(\ref{e_general}) and (\ref{e_final}) in this interval, instead of 0--4~kpc. This provides a
measurement of the surface density of the mass enclosed between 1.5 and 4~kpc from the plane only. This
approach has the advantage of being free of the uncertainties related to the total amount of visible mass and the
extrapolation of the kinematics to $Z\leq$1.5~kpc, although we have already demonstrated that these points
do not invalidate our conclusions. In fact, all the ISM and the majority of the stellar mass are found below
the range of integration, and a decrease of $\Sigma_\mathrm{disk}$(1.1~kpc) by 5~M$_\odot$~pc$^{-2}$ decreases the
expectation of $\Sigma_{>1.5}(Z)$ by only 0.15~M$_\odot$~pc$^{-2}$.

The derived profile of $\Sigma_{>1.5}(Z)$ is shown in the lower panel of Figure~\ref{f_resgen}. The same conclusions
as before can be drawn: the results perfectly match the expectation of the visible mass alone, and the estimated
local DM density is $\rho_{\odot,DM}=0.0\pm0.7$~mM$_\odot$~pc$^{-3}$. Moreover, the discrepancy with the models
comprising a DM halo is even more striking, since the OM model is 9$\sigma$ higher than the derived solution,
and the SHM and MIN models are excluded at the $\sim 7.5\sigma$ and $\sim 5\sigma$ level, respectively.


\section{ANALYSIS}
\label{s_anal}

The calculation relies on a set of kinematical measurements, ten hypotheses, and three parameters. In this section,
we will assess in more detail the reliability and robustness of the results, investigating if any of these input
quantities can be the cause of the mismatch between observations and model expectations, and if the DM halo could
pass undetected with our methods. Finally, alternative DM spatial distributions (non-sperical halo, dark disk, dark
ring) will also be considered and compared to the observations.

\subsection{The incidence of the parameters}
\label{s_an_param}

\begin{figure*}
\epsscale{1.}
\plotone{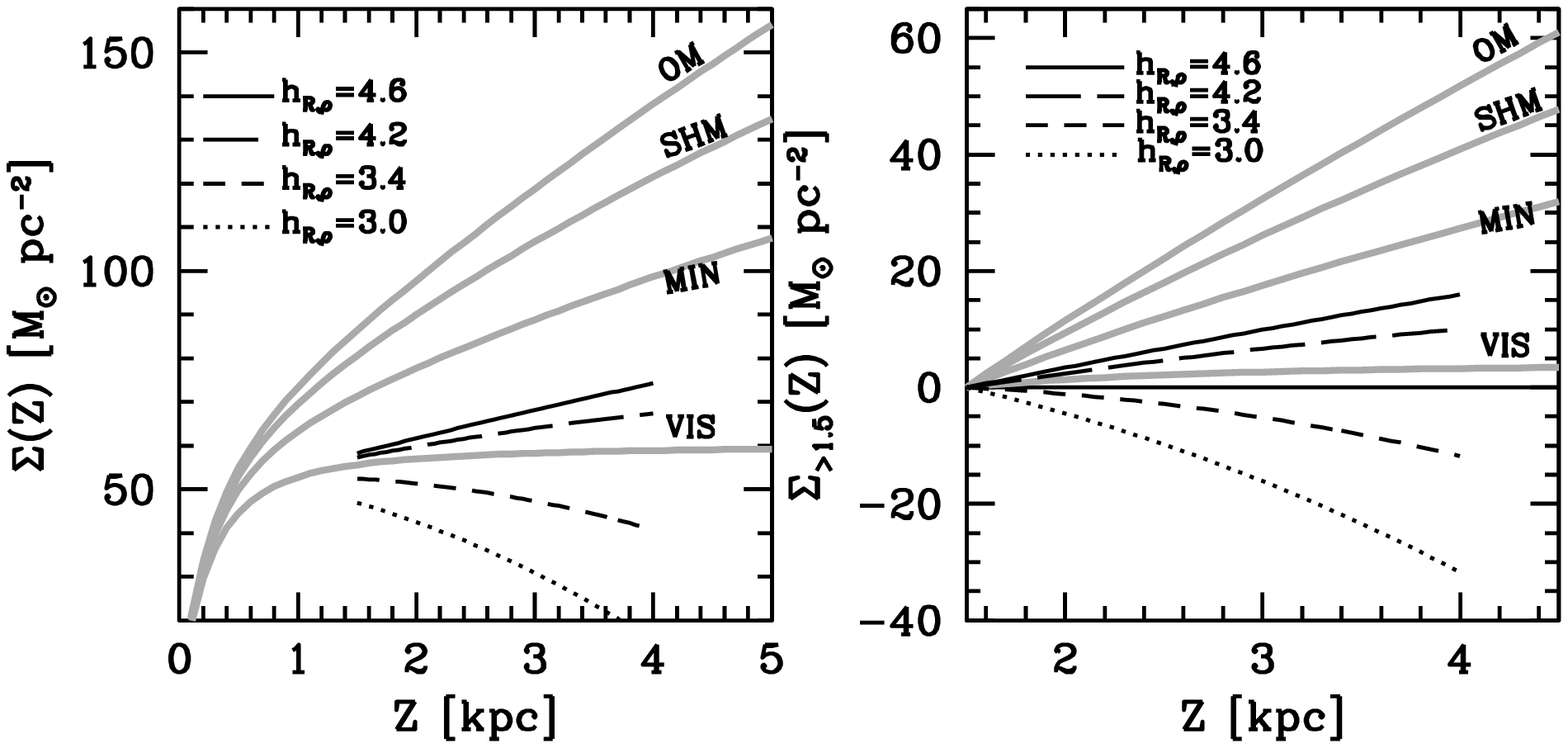}
\plotone{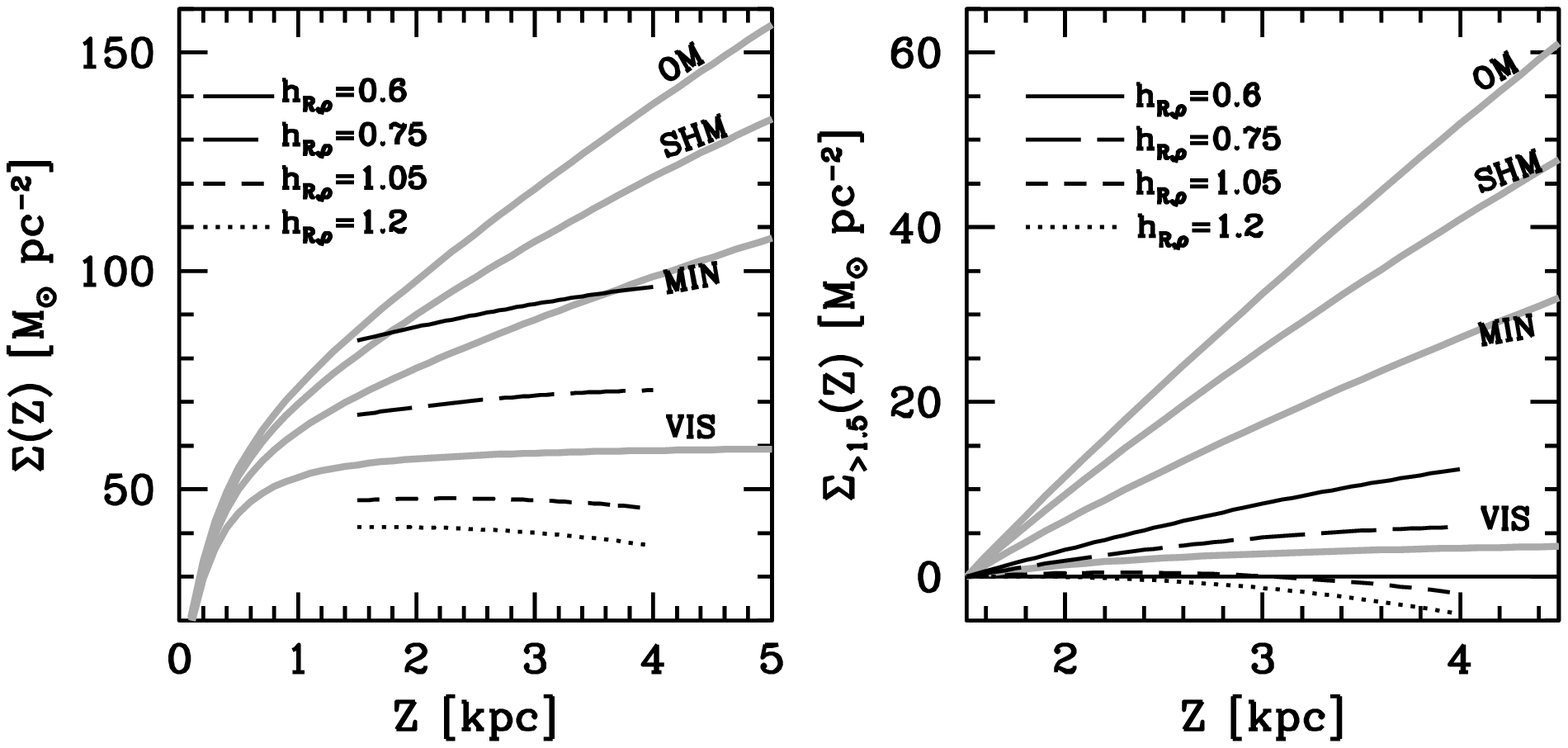}
\caption{Absolute (left panels) and incremental (right panels) surface density calculated with different values of the
thick disk scale length (upper panels) and scale height (lower panels), overplotted to the expectations of the models
described in the text (grey curves).
\label{f_param}}
\end{figure*}

A different definition of the solar Galactocentric distance has a negligible impact on the results. Both $\Sigma (Z)$
and $\Sigma_{>1.5}$(4~kpc) decrease with $R_\odot$, but they change by less than $\pm 3.5$~M$_\odot$~pc$^{-2}$ and
$\pm 2.6$~M$_\odot$~pc$^{-2}$ ($\sim 0.5\sigma$ at 4~kpc), respectively, when $R_\odot$ is varied in the range
7.5--8.5~kpc. The precise value adopted for $R_\odot$ is therefore irrelevant.

The effects of varying the thick disk scale height and length in the range 0.6--1.2~kpc and 3.0--4.6~kpc,
respectively, are shown in Figure~\ref{f_param}. Decreasing $h_{Z,\rho}$ shifts the curve of $\Sigma (Z)$ to higher
values, but does not affect its slope noticeably. For example, $\Sigma$ (1.5~kpc) agrees well with the expectation of
the most massive DM halo model (OM) when assuming $h_{Z,\rho}$=0.6~kpc, but at $Z$=4~kpc they differ by
$\sim 4\sigma$, because the gradient of the observed curve is too shallow. In fact, even assuming this extremely low
scale height, the less massive model (MIN) is still 2.5$\sigma$ higher than the curve of $\Sigma_{>1.5}(Z)$. On the
contrary, varying the thick disk scale length affects the slope of $\Sigma (Z)$, but does not add a lot of mass to the
derived solution. $\Sigma_{>1.5}(Z)$ increases with $h_{R,\rho}$, but even assuming $h_{R,\rho}$=5~kpc, the MIN model
is still 2$\sigma$ higher than both $\Sigma (Z)$ and $\Sigma_{>1.5}(Z)$.

\begin{figure}
\epsscale{1.}
\plotone{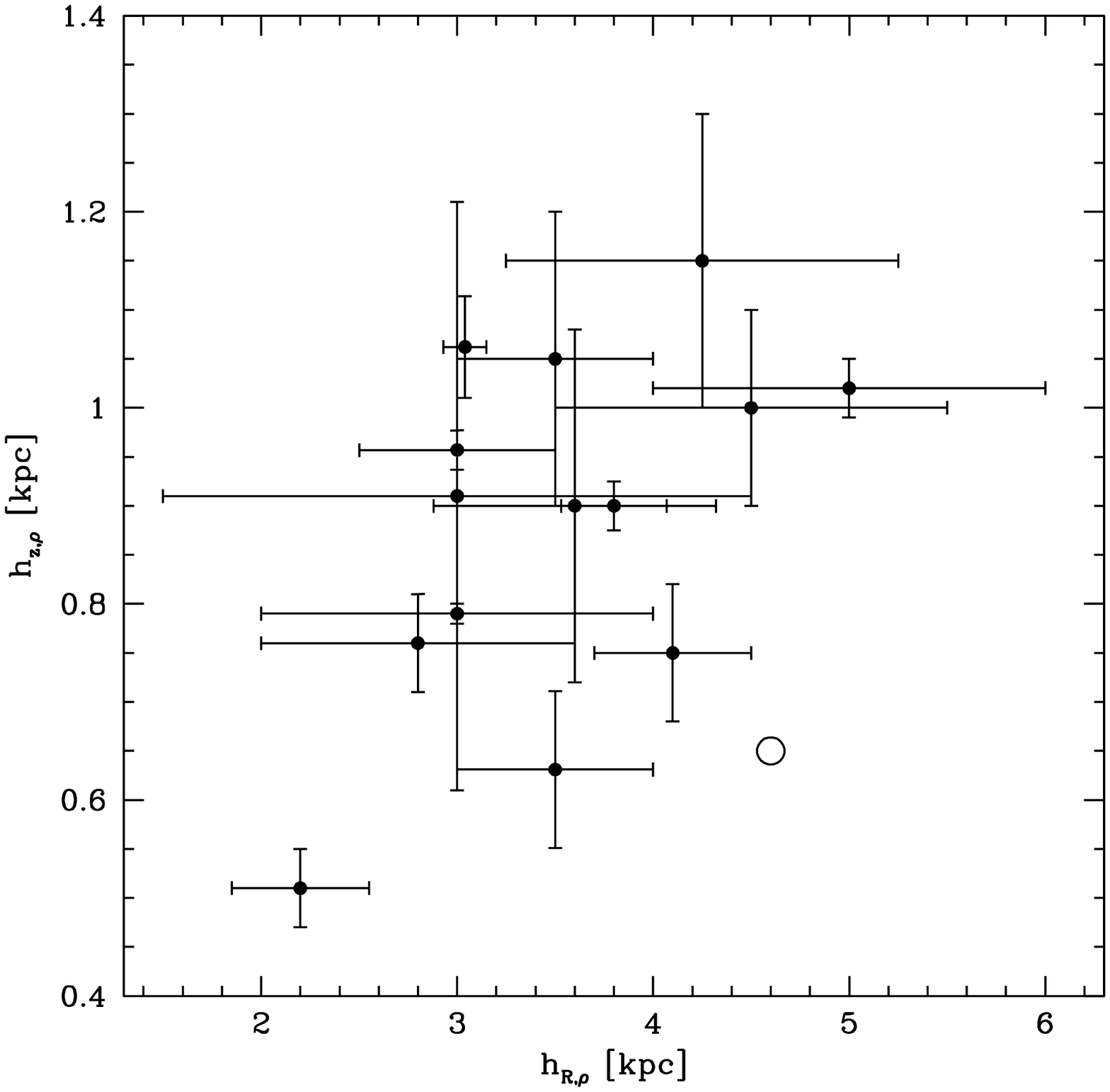}
\caption{Measured thick disk scale length and height in the
literature (full dots with 1$\sigma$-error bars). The empty circle shows the value required to force the calculated $\Sigma (Z)$
to match the MIN model.
\label{f_hRhz}}
\end{figure}

While varying $h_{Z,\rho}$ or $h_{R,\rho}$ alone is not enough to reconcile the measurements with the presence of a
classical DM halo, they can be changed simultaneously. Nevertheless, observational constraints prevent from
obtaining a solution overlapping the expectations of DM halo models, because even the less massive MIN model can be
roughly matched by the observations only by assuming $h_{Z,\rho}$=0.65~kpc and $h_{R,\rho}$=4.7~kpc. Such a large scale
length was proposed by some authors \citep[][]{Ratnatunga89,Beers95,Chiba00,Larsen03,Chang11}, but it was always
associated to a higher scale height, as
shown\footnote{The data of Figure~\ref{f_hRhz} are taken from:\citet{Yamagata92,Robin96,Ng97,Ojha99,Buser98,Buser99,
Siegel02,Cabrera05,Cabrera07,Bilir08,Juric08,deJong10,Carollo10,Chang11}} in Figure~\ref{f_hRhz}. The observations
indicate that the required thin and extended thick disk is very unlikely. In conclusion, varying the three parameters
involved in the calculations within the ranges allowed by the literature does not solve the problem of the missing DM
in the volume under analysis.

\subsection{Non-flat rotation curve}
\label{s_an_noflat}

The behavior of the Galactic rotation curve is still debated for $R\geq$11~kpc \citep[e.g.,][]{Binney97}, but there
is a general consensus that it is flat at the solar Galactocentric position. Indeed, recent observations find
only a tiny negative gradient
\citep[][$\partial \overline{V}/\partial R=-0.85$ and $-0.006\pm 0.016$~km~s$^{-1}$~kpc$^{-1}$, respectively]{Xue08,Fuchs09}.
Large deviations from a flat curve are therefore excluded, and theoretical models predict
$\vert\partial \overline{V}/\partial R\vert \leq 6$~km~s$^{-1}$~kpc$^{-1}$ \citep[e.g.,][]{Dehnen98,Olling00}.
\citet{Levine06} also used a flat curve to measure the Galactic density distribution.

If the hypothesis~(\ref{h_flat}) is dropped and a non-flat rotation curve is considered, the additional term
\begin{equation}
\frac{1}{\pi GR}\int_{0}^{Z}\frac{\partial \overline{V}}{\partial R}\overline{V} dz
\label{e_noflat}
\end{equation}
is added to the right hand term of both Equation~\ref{e_general} and~\ref{e_final}. In the following, we will
assume $V_\mathrm{LSR}$=220~km~s$^{-1}$, and the vertical shear of the thick disk rotational velocity
$\overline{V}(Z)$ will be represented by the linear expression of \citet{Moni12}, that well approximates the
underlying low-index power law in the $Z$-range under analysis. It is immediately evident from
Equation~\ref{e_noflat} that a decreasing rotation curve at $R_\odot$
($\frac{\partial \overline{V}}{\partial R}\leq 0$~km~s$^{-1}$~kpc$^{-1}$) introduces a negative term to the
calculation, thus increasing the discrepancy between the observations and the expectations of the DM halo models.
On the contrary, a higher $\Sigma (Z)$ is obtained if the rotation velocity increases with $R$, and a certain
amount of DM is thus allowed in the volume under analysis. However,
$\frac{\partial \overline{V}}{\partial R}= 10$~km~s$^{-1}$~kpc$^{-1}$ is required to match the minimum DM density
deduced by the Galactic rotation curve (MIN model), and
$\frac{\partial \overline{V}}{\partial R}= 16.5$~km~s$^{-1}$~kpc$^{-1}$ to match the SHM. Such steep rotation
curves are excluded by observations. Moreover, we are comparing the observations with models whose density was
fixed to return a flat rotation curve, hence we step into the contradiction that, while assuming a steeper curve
to increase the measured mass, we simultaneously increase the amount of DM required by the models. Among
the possibilities offered by the literature, the solution closest to the expectations of a DM halo model is
obtained assuming a rotation curve increasing its steepness from
$\frac{\partial \overline{V}}{\partial R}\Bigl{\vert_{Z=0}}=0$ to
$\frac{\partial \overline{V}}{\partial R}\Bigl{\vert_{Z=4 \mathrm{kpc}}}\approx 7$~km~s$^{-1}$~kpc$^{-1}$, as
modeled by \citet{Kalberla07} for the Galactic gas. Even in this case, however, the MIN model is 2$\sigma$ higher
than both the measured $\Sigma_{>1.5}(Z)$ and $\Sigma$ (4~kpc), while the SHM is excluded at the 5$\sigma$ level.
In conclusion, assuming a non-flat rotation curve can alter the results, but insufficiently to justify the
mismatch between the observations and the models.

\subsection{Alternative radial profiles of the dispersions}
\label{s_an_rad}

\begin{figure}
\epsscale{1.}
\plotone{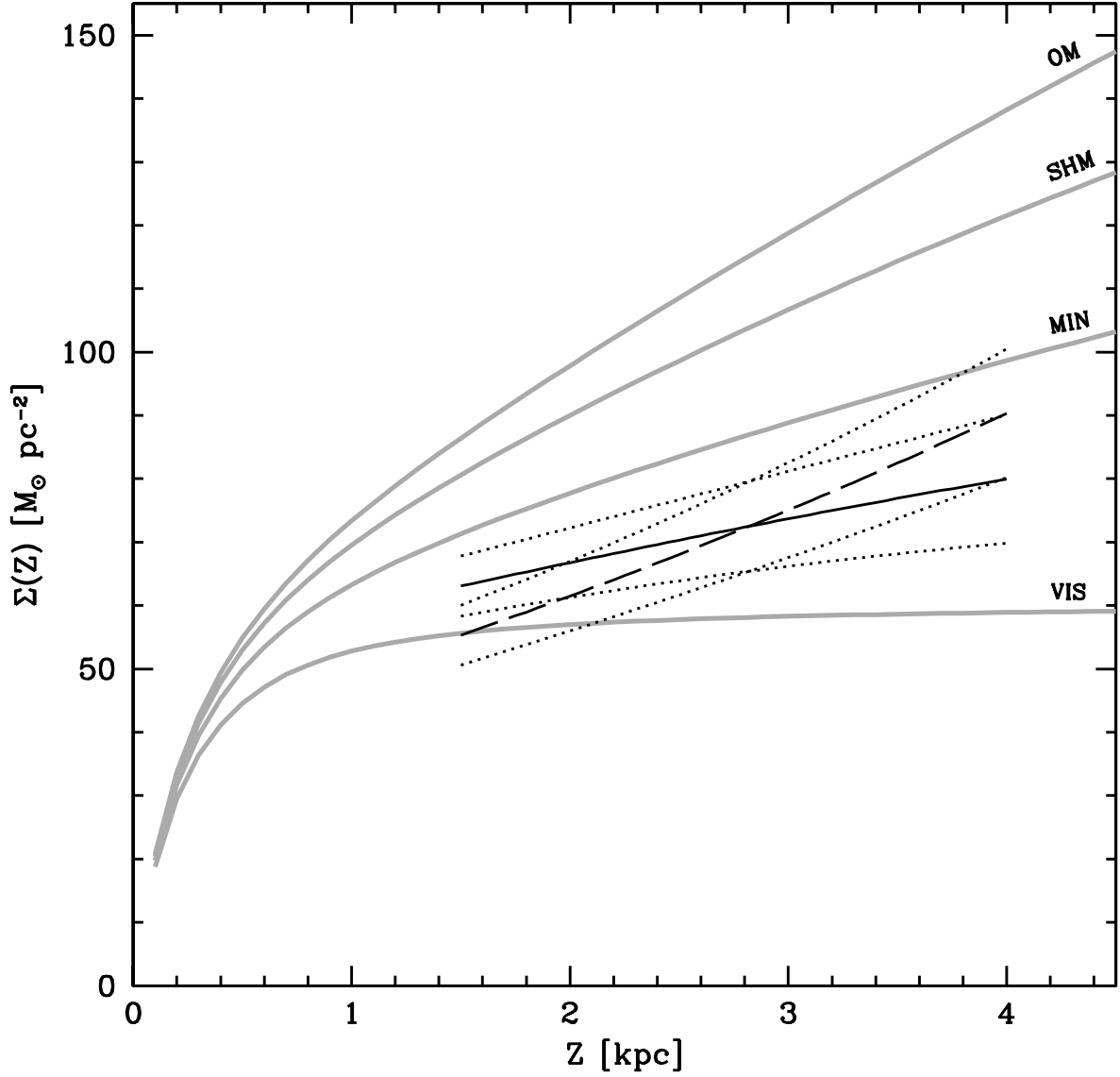}
\caption{Surface density as a function of Galactic height calculated assuming the radial constancy of the Toomre
Q-parameter (full black curve), and a linear radial decay of the dispersion (dashed curve). The dotted curves
indicate the respective 1$\sigma$ strip. The models discussed in the text are also overplotted (grey lines).
\label{f_toomre}}
\end{figure}

A popular model for the radial dependence of $\sigma_\mathrm{U}$, alternative to the hypothesis~(\ref{h_radial}),
is the assumption of a constant Toomre Q-parameter. Its validity is controversial, and it is disfavored by the
numerical integrations of \citet{Cuddeford92}. If the rotation curve is flat, the constant-Q model predicts that
\begin{equation}
\sigma^2_\mathrm{U}(R)\propto R^2 \exp\Bigl{(}-\frac{2R}{h_{R,\rho}}\Bigl{)}
\label{e_toomre}
\end{equation}
\citep{Amendt91}, from which it immediately follows that
\begin{equation}
\frac{\partial \sigma^2_\mathrm{U}}{\partial R}=\sigma^2_\mathrm{U}\cdot \Bigl{(}\frac{2}{R}-\frac{2}{h_{R,\rho}}\Bigl{)}.
\label{e_toomreder}
\end{equation}
When dropping the hypothesis~(\ref{h_radial}) to assume the constant-Q model, a choice about the radial
dependence of $\sigma^2_\mathrm{V}$ and $\overline{UW}$ must also be done. However, they have a lower incidence
on the results, and from Equation~\ref{e_toomreder} it is easy to see that
$\partial \sigma^2_\mathrm{U}/\partial R$ is almost indistinguishable in the two cases at the solar position,
because $R_\odot \approx 2h_{R,\rho}$. Hence, it makes little difference if they follow the trend of
Equation~\ref{e_toomreder}, or a purely exponential decay. This is not the same for $\sigma^2_\mathrm{U}$,
because the second derivative $\partial^2 \sigma^2_\mathrm{U}/\partial R^2$ enters into Equation~\ref{e_general}.

Substituting the assumption~(\ref{h_radial}) with Equation~\ref{e_toomreder} for $\sigma^2_\mathrm{U}$ leads
to a solution identical to Equation~\ref{e_final}, with $k_1$ replaced by
\begin{equation}
k_1\arcmin=\frac{15}{R_\odot\cdot h_{R,\rho}}-\frac{6}{R_\odot^2}-\frac{6}{h_{R,\rho}^2}.
\label{e_ktoomre}
\end{equation}
The results are shown in Figure~\ref{f_toomre}. The curve of $\Sigma (Z)$ both increases its slope and shifts
upward, and a certain amount of DM is allowed, but this is still insufficient when compared to the models:
the expectation of the SHM is still $\approx 4.5\sigma$ higher than both $\Sigma (Z)$ and $\Sigma_{>1.5}(Z)$,
while the discrepancy reduces to 2$\sigma$ for the minimum-density MIN model. The assumption of the constant-Q
model for the radial decay of $\sigma^2_\mathrm{U}$ mitigates the gap between the observations and the models,
but not enough to reconcile them.

\citet{Neese88} found that their measurements of $\sigma_\mathrm{U}$ throughout the Galactic disk could be
fitted with a linear radial decay, with slope $-3.8\pm 0.6$~km~s$^{-1}$~kpc$^{-1}$. The effects of dropping the
assumption~(\ref{h_radial}) in favor of linear radial profiles can be explored, although there is no theoretical
support for this behavior. We will assume $\frac{\partial \sigma_\mathrm{U}}{\partial R}$ from \citet{Neese88},
and $\frac{\partial \sigma_\mathrm{V}}{\partial R}=-6.0$~km~s$^{-1}$~kpc$^{-1}$ from \citet{Dinescu11}, whose
results for $\sigma_\mathrm{U}$ are identical to those of \citet{Neese88}, although they do not claim that the
underlying trend should be linear. Fixing $\frac{\partial \overline{UW}}{\partial R}$ is more problematic
due to the absence of measurements in the literature, but from the last two terms of Equation~\ref{e_general}
it is easy to see that a more negative value decreases the slope of $\Sigma (Z)$. Therefore, we will adopt
the most favorable case $\overline{UW} (R)$=constant.

Assuming the radial linear trends given above, the calculation of $\Sigma (Z)$ from Equation~\ref{e_general}
is straightforward, and the results are shown in Figure~\ref{f_toomre}. The curve of $\Sigma (Z)$ has a much
steeper slope than under the assumption~(\ref{h_radial}), and $\Sigma_{>1.5}(Z)$ matches the expectation of
the MIN model, agreeing with the SHM model within $\sim 1\sigma$. However, still $\Sigma (Z)$ is offset by
10--15~M$_\odot$~pc$^{-2}$ in the whole $Z$-range when compared to the MIN model, and
25--30~M$_\odot$~pc$^{-2}$ with respect to the SHM curve. Thus, the theoretically unjustified linear radial
profile of $\sigma_\mathrm{U}$ still requires an additional ad-hoc correction, like a strong overestimate
of the visible mass, or a thick disk scale height of $h_{Z,\rho}$=0.7~kpc, and it is therefore unlikely.

\subsection{A flared thick disk}
\label{s_an_flare}

\begin{figure}
\epsscale{1.}
\plotone{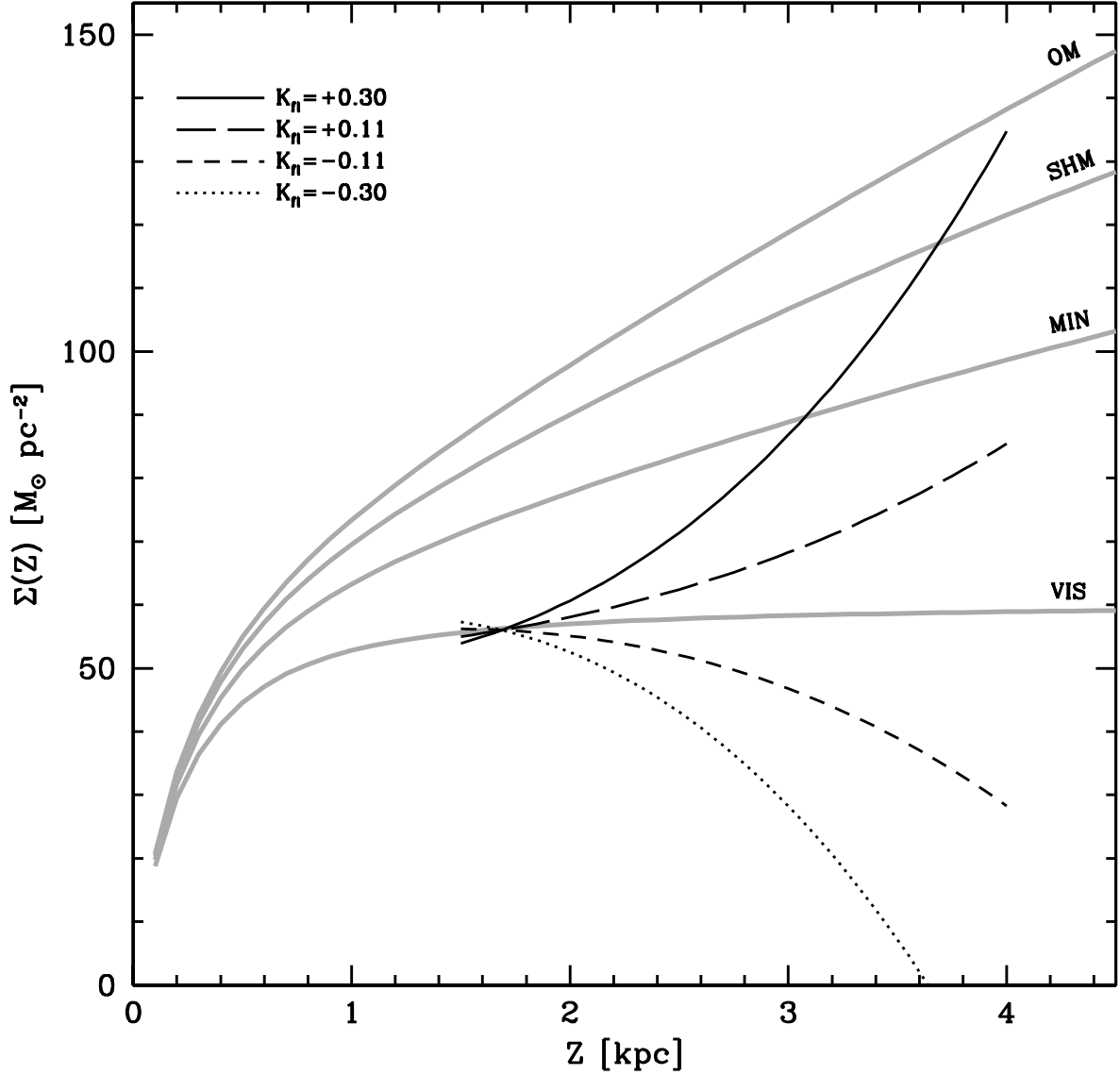}
\caption{Surface density as a function of scale height when a flared disk is considered (black lines), for
different values of the local flaring K$_{fl}=\frac{\partial \mathrm{h}_{Z,\rho}}{\partial R}$. The models
described in the text are also indicated with grey curves.\label{f_flare}}
\end{figure}

The flare of the Galactic thick disk is still an issue of debate, and the observations have not provided
unique evidence of its existence. \citet{Siegel02} and \citet{Du06}, for example, argue in favor of
a flared thick disk with scale height increasing with $R$, while \citet{Cabrera07}, supported by \citet{Bilir08}
and \citet{Yaz09}, proposed a flare of opposite sign. Given these contradictory results, it is most probably safe
to assume in our calculations that the flare is negligible at the solar position, else a relevant variation of
the thick disk scale height would have been detected beyond doubt.

If we assume a flared thick disk, thus dropping the assumption~(\ref{h_flare}), the scale height in
Equation~\ref{e_expo} becomes a function of $R$, and we have:
\begin{equation}
\frac{\partial \ln{(\rho)}}{\partial R}=-\frac{1}{h_{R,flare}}=
\Bigl{(}-\frac{1}{h_{R,\rho}}+\frac{Z}{h^2_{Z,\rho}(R)}\cdot\frac{\partial h_{Z,\rho}}{\partial R}\Bigl{)}.
\label{e_flare}
\end{equation}
The consequence of this new approach is the simple substitution of $h_{R,\rho}$ with $h_{R,flare}$ in
Equation~\ref{e_general}, plus the additional term
\begin{equation}
-\int_{0}^{Z}\sigma^2_U \cdot\frac{\partial}{\partial R}\Bigl{(}\frac{1}{h_{R,flare}}\Bigl{)}dz.
\label{e_flare2}
\end{equation}
The new formulation, although analytically simple, hides many practical complexities: first, the expression
for $h_{Z,\rho} (R)$ must be accurate enough to provide a good approximation up to its second radial
derivative. Additionally, the assumption~(\ref{h_radial}) implicitly relies on the constancy of
$h_{Z,\rho} (R)$ \citep{Kruit81,Kruit82}, and in a flared disk the radial behavior of the dispersions is harder
to model. Here we will therefore limit to the first-order approximation that the flare is locally small, hence
well represented by a linear function of $R$ \citep[as proposed by][]{Cabrera07}, and not noticeably
affecting the radial behavior of the dispersions. In this case, the additional term of Equation~\ref{e_flare2}
is neglected, and a solution identical to Equation~\ref{e_final} is obtained, with $k_1$ and $k_3$
respectively substituted by
\begin{eqnarray}
&&k_1\arcmin=\frac{1}{R_\odot\cdot h_{R,flare}}-\frac{1}{h_{R,\rho}^2}+
\frac{1}{h_{R,\rho}}\cdot\Bigl{(}\frac{2}{R_\odot}-\frac{1}{h_{R,flare}}\Bigl{)}, \\
&&k_3\arcmin=\frac{2}{h_{R,\rho}}+\frac{1}{h_{R,flare}}-\frac{2}{R_\odot}.
\end{eqnarray}
The resulting profiles of the surface density are shown in Figure~\ref{f_flare}. The introduction of a flare
in the formulation only changes the slope of the curve, that decreases in the case of a negative flare
($\frac{\partial h_{Z,\rho}}{\partial R}<0$). A positive flare, however, does not return results that can be
easily reconciled with the presence of a DM halo: $\Sigma_{>1.5}(Z)$ can be forced to match the MIN model
assuming $\frac{\partial h_{Z,\rho}}{\partial R}=0.1$, and the SHM model with
$\frac{\partial h_{Z,\rho}}{\partial R}=0.16$, but even in this case $\Sigma (Z)$ still remains
15--20~M$_\odot$~pc$^{-2}$ lower than the expectations of this model in the whole $z$ range. As in the case
of a linear radial decay of the dispersion (Section~\ref{s_an_rad}), an additional correction to the
formulation is therefore required.

\subsection{Alternative kinematical results}
\label{s_an_dinescu}

\begin{figure}
\epsscale{1.}
\plotone{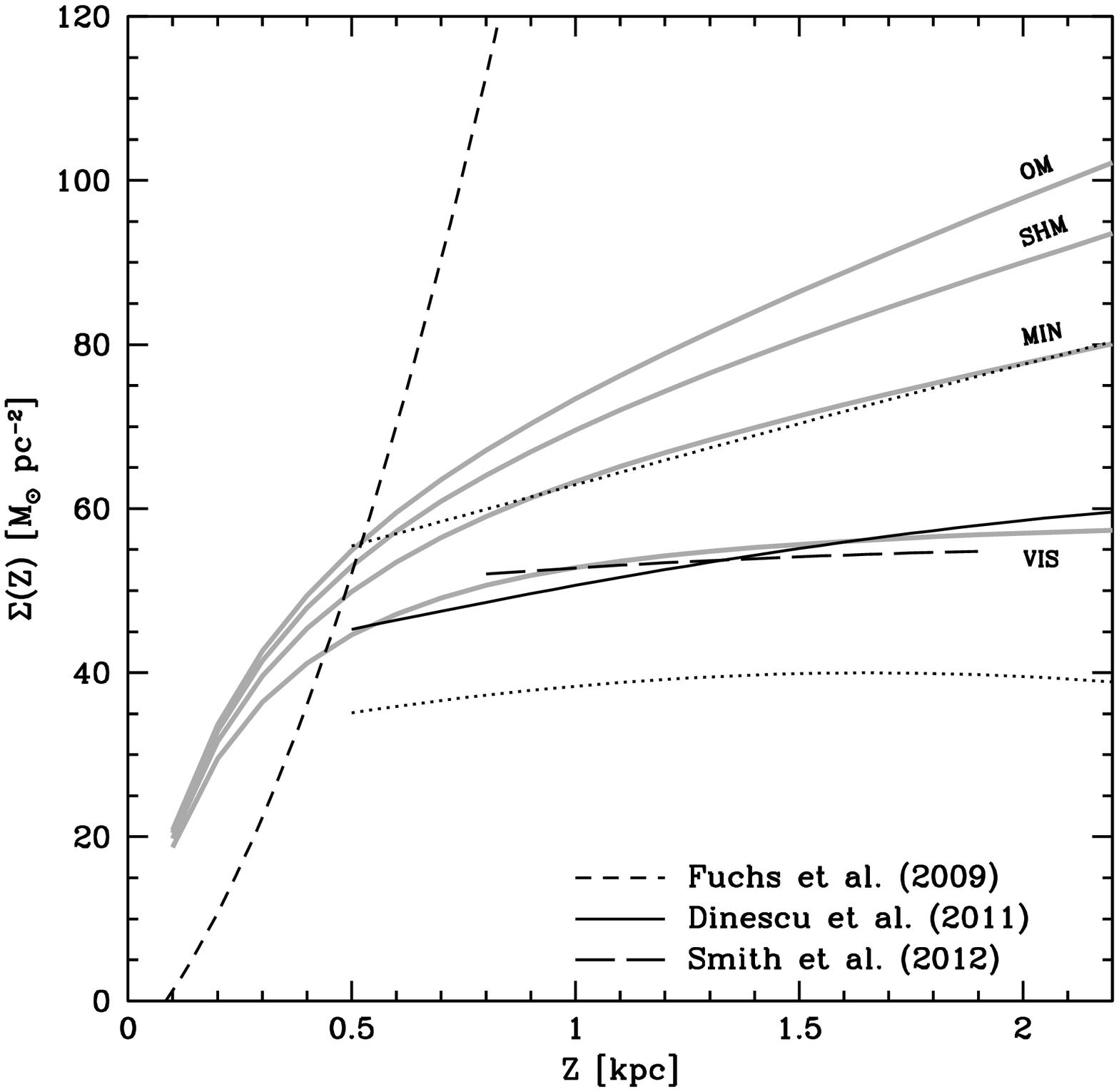}
\caption{Surface density calculated by use of the kinematical results of \citet[][short-dashed curve]{Fuchs09},
\citet[][full curve with dotted 1$\sigma$ stripe]{Dinescu11}, and \citet[][long-dashed curve]{Smith12}. The
expectations of the models discussed in the text are overplotted.
\label{f_din}}
\end{figure}

We repeated the calculations replacing the kinematical measurements of \citet{Moni12} with other works in the
literature, to check the incidence of the assumed kinematics on our results. Three previous
investigations are suitable for our purposes: \citet{Fuchs09}, \citet{Dinescu11}, and \citet{Smith12}. The
results are shown in Figure~\ref{f_din}.

The data points of \citet{Fuchs09} in the range $Z$=0--0.8~kpc were linearly fitted to derive the vertical
profiles of $\sigma_\mathrm{U}$, $\sigma_\mathrm{V}$, $\sigma_\mathrm{W}$, and $\overline{UW}$ to be inserted
in the Equation~(\ref{e_final}). A quadratic fit led to negligible differences. We assumed $h_{Z,\rho}$=0.3~kpc
and $h_{R,\rho}$=2.6~kpc \citep{Juric08}, because their sample mostly comprises thin disk stars. The curve of
$\Sigma (Z)$ thus derived is clearly unphysical. However, \citet{Binney09} have already shown that the
kinematical data of \citet{Fuchs09} lead to inconsistent results when used to constrain the mass distribution
in the Galaxy. The most likely cause of the problem is that their sample is a mixture of old thin and thick
disk stars, with the incidence of the thick disk increasing with $Z$, causing too steep a vertical gradient of
the dispersions. On the contrary, \citet{Dinescu11} used a pure thick disk sample to measure the variation of
the kinematics in the range $Z$=0.3--2.2~kpc. Unfortunately, they did not give the $Z$-profile of
$\overline{UW}$, but their measurement of the tilt angle $\alpha$ (strongly related to  $\overline{UW}$) agrees
with \citet{Moni12}. Hence we assumed the same profile of $\overline{UW} (Z)$ as before. In any case, the
incidence of this term is very limited at lower Galactic heights. The use of the \citet{Dinescu11} data
returns the same general results previously discussed: $\Sigma (Z)$ well matches the expectations for the
visible mass alone even in this case. Unfortunately, the results are weaker, both because of the larger
observational errors, and because the expectations of the models are less distinct at lower $Z$. Thus, the less
massive DM model is only 1$\sigma$ higher than the calculated solution.

\citet{Smith12} recently measured the change of the three-dimensional disk kinematics between $Z$=0.5 and
1.8~kpc, in three metallicity bins. The metal-rich group ([Fe/H]$\geq -0.5$) is probably a heterogeneous mix of
thin disk sub-populations, and it is not suitable for our purposes. The metal-poor thick disk, dominating the
sample with [Fe/H]$\leq -0.8$, has a unique spatial distribution \citep{Carollo10} poorly studied in the
literature. Hence, we adopted the results for the intermediate-metallicity stars ($-0.8\leq$[Fe/H]$\leq -0.5$),
representative of a thick disk stellar population similar to that studied by \citet{Moni12}. To derive the
vertical trend of the kinematical quantities, we linearly fitted their results excluding their nearest bin
($\overline{Z}$=0.7~kpc). We thus limited the incidence of a residual thin disk contamination, not efficiently
excluded at low $Z$ by the metallicity-only sample selection. Moreover, restricting the $Z$-range ensures that
the linear fit is a good approximation of the underlying trend because close to the plane, as already discussed
in Section~\ref{s_res}, the vertical gradient of the dispersions is expected to vary rapidly with $Z$. We did
not estimated an error on $\Sigma(Z)$ in this case, because the uncertainties associated to the kinematical
quantities are not well defined in our linear fit of three data points. As shown in Figure~\ref{f_din}, the
solution obtained with \citet{Smith12} overlaps that derived by means of the results of \citet{Dinescu11},
again roughly matching the expectations for the visible mass only. In conclusion, two more sets of
independent kinematical results return results identical to ours when used in the calculation of $\Sigma (Z)$,
although with a lower significance.

\subsection{DM halo detectability}
\label{s_an_detect}

\begin{figure}
\epsscale{1.}
\plotone{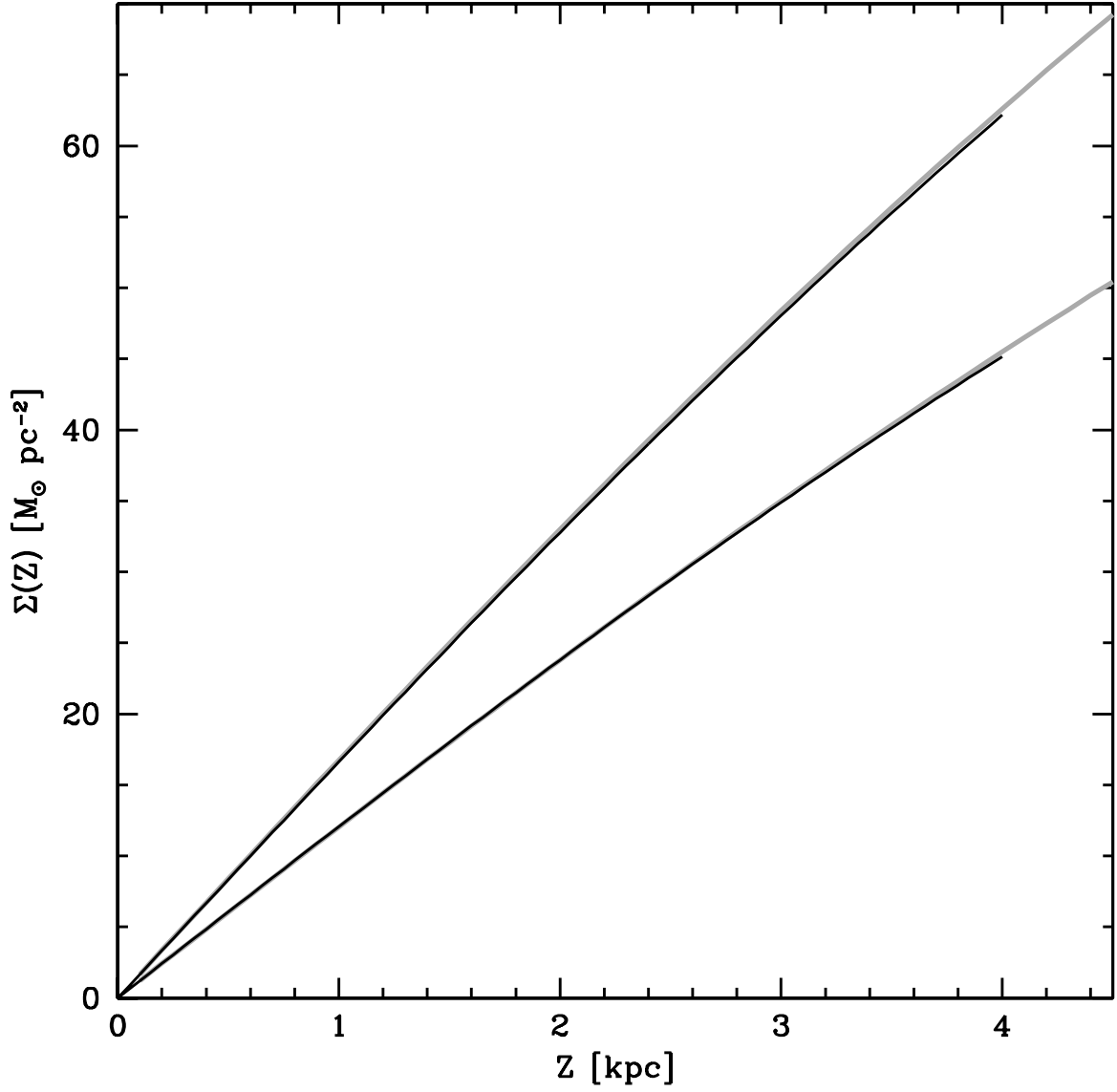}
\caption{Comparison between the surface density of the dynamical mass estimated by means of the Poisson equation
(black curves), and of the mass enclosed between $\pm Z$ (grey curves), for the DM halo models SHM and N97.
\label{f_sigdm}}
\end{figure}

The failure to detect the presence of a DM halo naturally raises the question whether this feature is detectable
by our method. In fact, the calculation does not measure the mass directly, but a variation of the gravitational
potential, from which the mass density is derived. \citet{Moni10} noted that, if the DM halo was too uniform,
extended, and poorly concentrated, it could cause a negligible change of the potential in the volume under
analysis, thus resulting undetectable.

The gravitational potential can be separated at any point as the sum of the potential of the dark and visible
matter, $\phi =\phi_\mathrm{DM}+\phi_\mathrm{VIS}$. It is easy to see that the dynamical surface density
calculated through Equation~(\ref{e_Pois}) can therefore be expressed as the sum of two contribution,
$\Sigma (Z)=\Sigma_{\phi,\mathrm{DM}}(Z)+\Sigma_{\phi,\mathrm{VIS}}(Z)$, where
\begin{equation}
\Sigma_{\phi,\mathrm{DM}}(Z)=\frac{1}{2\pi G}
\Bigl{[}\int_{0}^{Z} \frac{1}{R}\frac{\partial}{\partial R}\Bigl{(}R \frac{\partial \phi_\mathrm{DM}}{\partial R}\Bigl{)}dz
+\frac{\partial \phi_\mathrm{DM}}{\partial Z}\Bigl{]},
\label{e_PoisDM}
\end{equation}
and analogously for $\Sigma_{\phi,\mathrm{VIS}}(Z)$. The potential of a spherical NFW DM distribution can be
written as
\begin{equation}
\phi_\mathrm{DM}(s)=-g(c)\frac{GM_\mathrm{V}}{r_\mathrm{s} c}\frac{\ln{(1+cs)}}{s},
\label{e_NFWpot}
\end{equation}
where $r_\mathrm{s}$ is a characteristic radial scale-length, $c$ is the concentration parameter
\citep{Lokas01}, $s$=$r/(r_\mathrm{s}c)$ with $r=\sqrt{R^{2}+Z^{2}}$ the radial spherical coordinate,
M$_\mathrm{V}$ is the total mass within the Viral radius
$r_\mathrm{V}=r_\mathrm{s}c$, and $g(c)=(\ln{(1+c)}-c/(1+c))^{-1}$. The two NFW models analyzed in this paper
have M$_\mathrm{V}=5.2\times 10^{11}$~M$_\odot$ and $c=16.8$ (SHM model), and
M$_\mathrm{V}=6.85\times 10^{11}$~M$_\odot$ and $c=9$ (N97 model). Inserting Equation~(\ref{e_NFWpot}) into
Equation~(\ref{e_PoisDM}), we can estimate $\Sigma_{\phi,\mathrm{DM}}(Z)$, i.e. the dynamical DM mass that
can be detected by means of the Poisson equation between $\pm Z$ from the Galactic plane. The results of this
calculation for the SHM and N97 models are shown in Figure~\ref{f_sigdm}. In both cases, the dynamical DM mass
perfectly matches the DM mass expected by the models, indicating that all the mass of the DM halo enclosed
between $\pm Z$ is detected by the use of Equation~(\ref{e_Pois}). It is therefore false that the small potential
difference induced by the DM halo between 0 and 4~kpc from the plane can induce an underestimate of the physical
mass in the volume under analysis.

\begin{figure}
\epsscale{1.}
\plotone{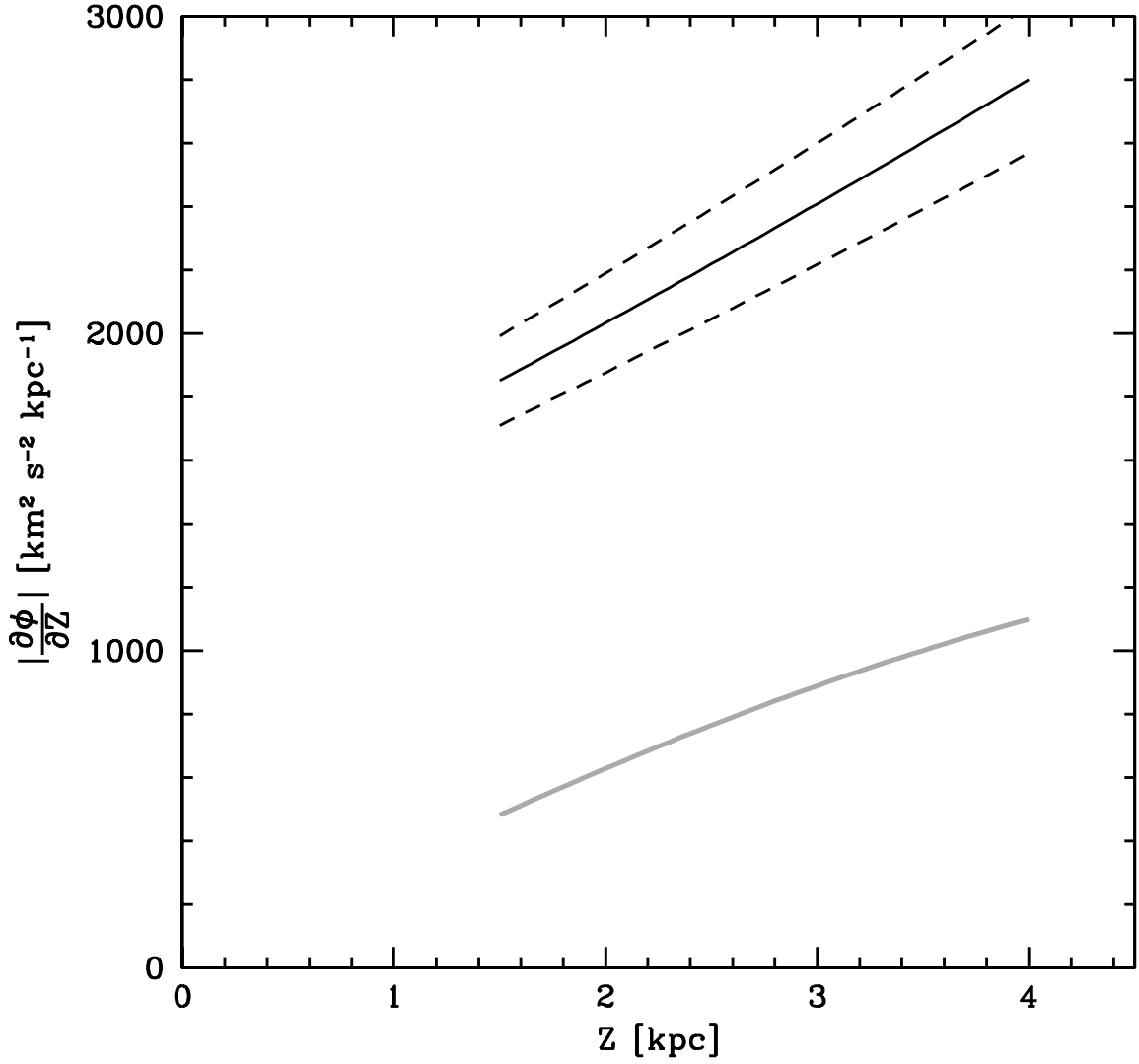}
\caption{The vertical gradient of the potential caused by the SHM dark halo model (grey curve), and measured 
from stellar kinematics through Equation~\ref{e_gradDM} (black solid curve), with its $\pm 1\sigma$ strip
(black dashed curves).
\label{f_phidm}}
\end{figure}

It can also be argued that, even if Equation~\ref{e_Pois} should not fail, the change of the stellar
kinematics between $Z$=1.5 and 4~kpc induced by the presence of a DM halo could be too small to be detected
by the kinematical measurements adopted in our work. To check this possibility, we compared the vertical
gradient of the potential of a DM halo, $\frac{\partial \phi_\mathrm{DM}}{\partial Z}$, with the observed
quantity
\begin{equation}
\frac{\partial \phi}{\partial Z}\Bigl{\vert}_\mathrm{obs}=\frac{\sigma^2_W}{h_{Z,\rho}}
-\frac{\partial \sigma^2_W}{\partial Z}
-\overline{UW}\cdot\Bigl{(}\frac{1}{R}-\frac{2}{h_{R,\rho}}\Bigl{)}.
\label{e_gradDM}
\end{equation}
This is the vertical gradient of the gravitational potential estimated by means of the stellar kinematics
through Equation~(\ref{e_Jeansz}). The comparison, relative to the SHM model, is shown in Figure~\ref{f_phidm}.
Clearly, the DM halo does not generate a strong vertical gradient of the potential: the surface density of the
expected DM mass enclosed within $\pm$4~kpc is similar to the measured one ($\sim$60~M$_\odot$~pc$^{-2}$),
while $\frac{\partial \phi_\mathrm{DM}}{\partial Z}$ is about a factor of two smaller than the observed
potential gradient. Nevertheless, $\frac{\partial \phi}{\partial Z}\Bigl{\vert}_\mathrm{obs}$ is measured with
an error of the order of 10\% ($\sim$200~km$^{2}$~s$^{-2}$~kpc$^{-1}$). This indicates that the presence of a
DM halo would had not passed undetected, because it would had caused a variation of the measured
$\frac{\partial \phi}{\partial Z}\Bigl{\vert}_\mathrm{obs}$ about 4--5 times larger than the observational
errors. In a forthcoming study, the orbit integration of a synthetic stellar population embedded in the
Galactic potential will be used to test the reliablility of our method, and to analyze in more detail the
causes of the systematics that can arise.

\subsection{Dark halo flattening}
\label{s_an_flattening}

\begin{figure*}
\epsscale{1.}
\plotone{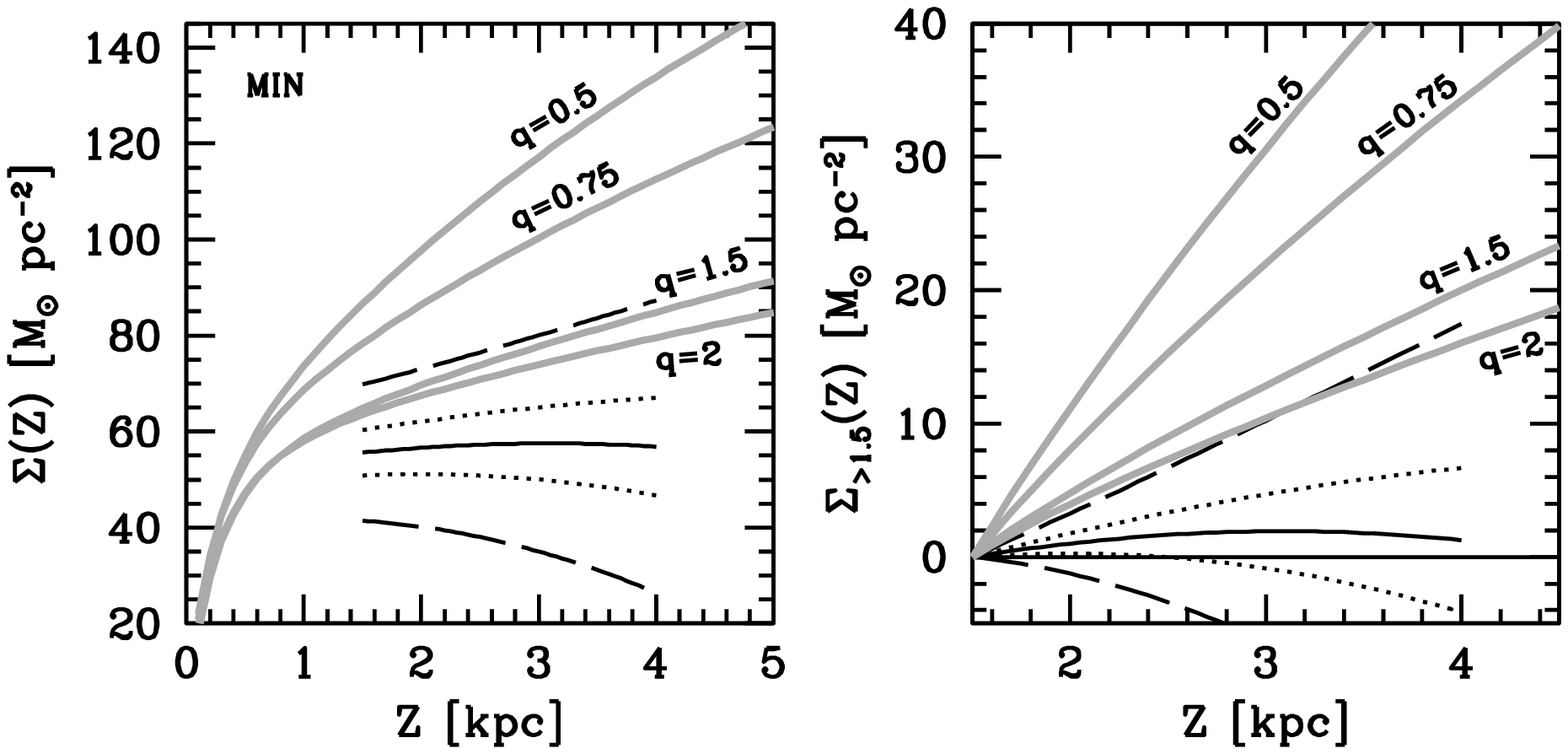}
\plotone{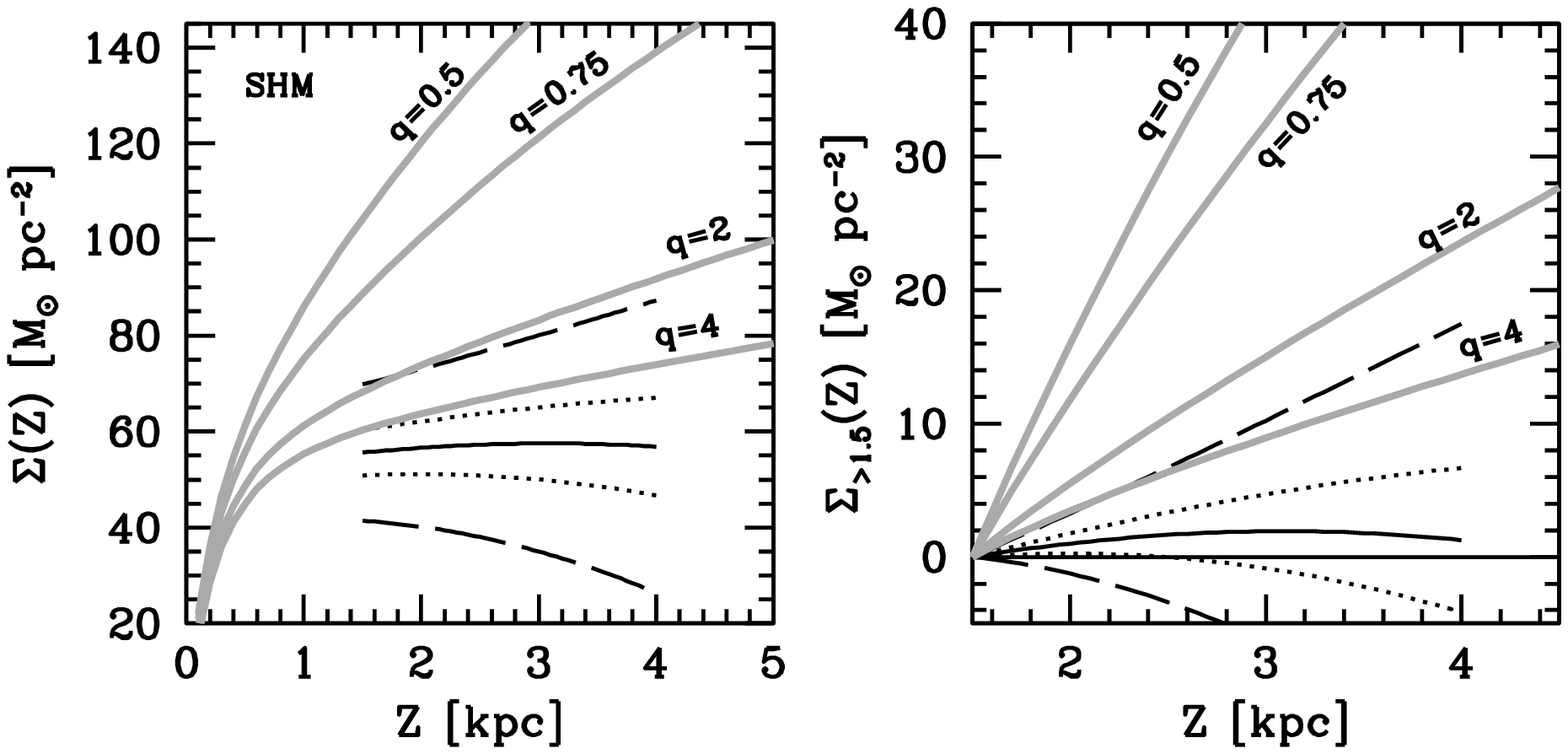}
\caption{Comparison between the measured surface density (black line) and the model expectations (grey
curves) with varying flattening parameter $q$. The dotted and dashed lines indicate the 1$\sigma$ and
3$\sigma$ stripes, respectively. Upper panels: MIN model. Lower panels: SHM model.
\label{f_flattening}}
\end{figure*}

So far, we have considered only spherical models for the DM halo, but the expected amount of DM in the
volume under analysis changes noticeably if its shape is modified. The effects of varying the flattening of
the models SHM and MIN are shown in Figure~\ref{f_flattening}. The models were calculated from
Equations~(\ref{eq_DMOll}) and (\ref{eq_DMgen}), where $R_c$, $\rho_c$, and $\rho_{\odot,DM}$ were
multiplied by 1/$q$ to keep the resulting Galactic rotation curve the same \citep{Olling95,Olling01}.
Clearly, oblate models ($q< 1$) imply a larger quantity of DM in the solar vicinity, and they depart
further from the observations. The expected surface density, on the other hand, decreases at increasing
$q$, and highly prolate models approach the observational results. A perfect match requires a zero-density
halo and is therefore never reached for any finite $q$, but a lower limit can be derived. We thus conclude
that strongly prolate models are required, because within 2$\sigma$ (95\% confidence level) we have
$q\geq 2$ for the MIN model, and this lower limit is much higher for more common, higher-density models,
for example $q\ga 4$ for the SHM.

\subsection{Other dark matter structures}
\label{s_an_disk}

\begin{figure*}
\epsscale{1.}
\plotone{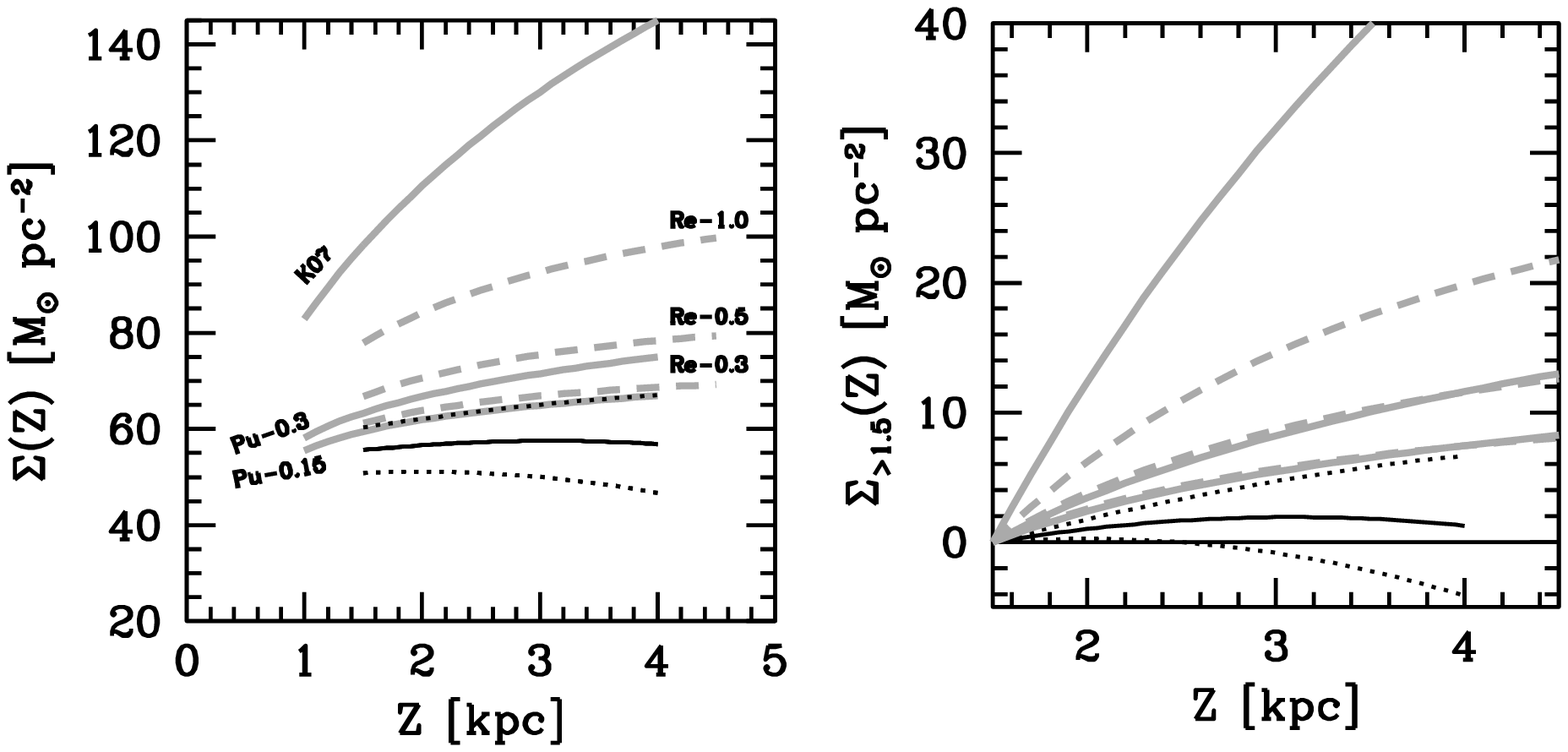}
\caption{Comparison between the observed surface density (black line), and the expectations of DM disk
models from the literature. K07: \citet{Kalberla03} model; Pu-0.3 and Pu-0.15: \citet{Purcell09} models
with local density $\rho_{\odot,DM}$=3 and 1.5~mM$_\odot$~pc$^{-3}$, respectively, and scale height
4.6~kpc; Re-1.0, Re-0.5, and Re-0.3: \citet{Read08} models with $\rho_{\odot,DM}$=10, 5, and
3~mM$_\odot$~pc$^{-3}$, respectively, and scale height 2.4~kpc.
\label{f_DMdisk}}
\end{figure*}

We have compared the observations with common models of DM halos, but other spatial distribution have been
presented in the literature, like a DM disk and a ring \citep[e.g.,][]{Kalberla07}. Nevertheless, these
features were usually proposed in addition and not as an alternative to a spheroidal halo, thus enhancing
the expected local DM density.

The presence of a DM ring at $R\gg R_\odot$ cannot be detected with our observations, but most of the mass
of a DM disk would be included in the volume under analysis, and thus detectable by our method.
The existence of such a feature,
first proposed by \citet{Lake89}, is of particular interest, because it is a natural expectation of
current $\Lambda$CDM models of Galactic formation \citep{Purcell09}. In Figure~\ref{f_DMdisk} our results are
compared to the expectations of the massive DM disk model of \citet{Kalberla03}, the thick, low-density
model of \citet{Purcell09}, and the thinner and denser ones from \citet{Read08}. The comparison shows that
the dark disk proposed as an alternative to a DM halo \citep{Kalberla03} is excluded by the observations with
very high significance (9$\sigma$) and, as discussed in the previous sections, this mismatch cannot be
corrected by simply altering one parameter or one assumption. On the contrary, the local density of the DM
disk models proposed by \citet{Read08} and \citet{Purcell09} is low, and they cannot be ruled out. Some
constraints can be derived, though. The curves of both $\Sigma (Z)$ and $\Sigma_{>1.5}(Z)$ indicate that,
within 2$\sigma$, the thick DM disk of \citet[][scale height $\sim$4.6~kpc]{Purcell09} must have a local
density $\rho_{\odot,DM}<3$~mM$_\odot$~pc$^{-3}$, while the upper limit for the thinner \citet{Read08} model
(scale height 2.4~kpc) is $\rho_{\odot,DM}<5$~mM$_\odot$~pc$^{-3}$. However it must be taken into
consideration that these results are sensitive to a change of a parameter or an assumption as
discussed in Sections~\ref{s_an_param} to~\ref{s_an_flare}. In any case, these DM disk models do not sustain
the Galactic rotation curve, and their presence in addition to a DM halo would enhance the discrepancy
between the observations and the expected DM density in the volume under study.


\section{DISCUSSION AND CONCLUSION}
\label{s_conc}

The measurement of the mass surface density at the solar Galactocentric position between 1.5 and 4~kpc from
the Galactic plane accounts for the visible mass only. The DM density in the solar neighborhood,
extrapolated from the observed curve of $\Sigma (Z)$, is $\rho_{\odot,DM}=0\pm1$~mM$_\odot$~pc$^{-3}$, at
variance with the general consensus that it must be in the range 5--13~mM$_\odot$~pc$^{-3}$
\citep[e.g.,][]{Weber10,Garbari11}. Our recent measurements of the thick disk kinematics were used in the
calculation, but the observed lack of DM is independent of this choice, because very similar results can be
obtained by means of other kinematical results in the literature. The calculation relies on three input
parameters and ten assumptions, but the observations cannot be reconciled with the DM halo modes modifying
one of them. Altering them at will introduces enough freedom to force the solution to match the expectations
of the most preferred model, but in this case an exotic series of unlikely hypotheses must be invoked. For
example, a very thin thick disk (h$_{Z,\rho}$=0.7~kpc), either very extended in the radial direction
(h$_{R,\rho}$=4.6~kpc) or strongly flared at the solar position ($\frac{\partial h_{Z,\rho}}{\partial R}$=0.1)
make the solution coincide with the minimum DM local density deduced from the Galactic rotation curve, but
the observational constraints exclude or disfavor these scenarios. On the other hand, the expected visible
mass strikingly matches the observations without any effort, by use of the most probable assumptions. This
coincidence lends weight to the interpretation of these results, because it is easy to obtain an unphysical
solution if one or more wrong hypotheses are being made. For example, \citet{Moni10} showed that the results
obtained under the assumption of a cross-term $\overline{UW}$ symmetric with respect to the Galactic plane, as
an alternative to our hypothesis~(\ref{h_uw}), violate two minimum requirements: the surface density must at least
account for the known visible matter, and it cannot decrease with $Z$. Thus, the excellent agreement between
the measured mass and the visible mass is unlikely to have been obtained by pure chance. We also demonstrated
that our method cannot fail to detect the presence of a classical DM halo, because it causes a noticeable
change in the stellar kinematics, one order of magnitude larger than the observational errors.

The only viable solution to reconcile the observations with the models is the assumption of a highly prolate
DM halo, that can sustain a flat rotation curve with a negligible density at the solar position.
Observational constraints on the DM halo shape are scarce and often controversial and, while spherical
structures are usually preferred \citep[e.g.][]{Ibata01,Majewski03,Johnston05,Fellhauer06}, a prolate spheroid
($q=5/3$) was invoked by \citet{Helmi04} and \citet{Law05}. Our observations suggest that $q\geq$2 is required
even in the lower bound case of the least massive model, for agreement within 2$\sigma$. However, very prolate
structures are atypical in cold dark matter simulations, which have problems in reproducing them
\citep[e.g.,][]{Dubinski91}. Thus, it must still be proven that a DM halo with a high flattening parameter is
fully compatible with the current $\Lambda$CDM paradigm.

A dark spheroidal component is required to sustain the Galactic rotation curve, observationally confirmed to
be flat for $R\geq$5~kpc \citep{Kalberla07,Sofue08,Xue08}. In the presence of the visible mass only, Newtonian
dynamics would predict a steep keplerian fall-off. Very noticeably, the calculation returns the unphysical
result of a total surface density decreasing with $Z$ if a rapidly decreasing curve is assumed. Hence,
a flat rotation curve is required, while failing to detect the DM necessary to sustain it. This apparent
contradiction is actually a confirmation that the calculation is reliable, because it is consistent with an
observationally proven fact, although not with its expected explanation.
 
In conclusion, the observations point to a lack of Galactic DM at the solar position, contrary to the
expectations of all the current models of Galactic mass distribution. A DM distribution very different to
what it is today accepted, such as a highly prolate DM halo, is required to reconcile the results with the DM
paradigm. The interpretation of these results is thus not straightforward. We believe that they require
further investigation and analysis, both on the observational and the theoretical side, to solve the problems
they present. We feel that any attempt to further interpret and explain our results, beyond that presented in
this paper, would be highly speculative at this stage. Future surveys, such as GAIA, will likely be crucial to
move beyond this point. However, as our results currently stand, we stress that, while numerous experiments seek
to directly detect the elusive DM particles, our observations suggest that their density may be negligible in
the solar neighborhood. This conclusion does not depend on the cause of this lack of DM at the solar position.
For example, if our results are interpreted as evidence of a highly prolate cold DM halo with $q\geq$2, this
would have a local density lower than 2~mM$_\odot$~pc$^{-3}$, i.e. more than a factor of four lower than what
usually assumed in the interpretation of the results of these experiments.

It is clear that the local surface density measured in our work, extrapolated to the rest of the Galaxy,
cannot retain the Sun in a circular orbit at a speed of $\sim$220~km~s$^{-1}$. A deep missing mass problem
is therefore evidenced by our observations. Indeed, we believe that our results do not solve any problem, but
pose important, new ones.


\acknowledgments
The authors warmly thank the referee C. Flynn for his extensive work of revision of our results, and for
his detailed criticisms. C.M.B. and R.A.M. acknowledge support from the Chilean Centro de Astrof\'isica FONDAP
No. 15010003, and the Chilean Centro de Excelencia en Astrof\'isica y Tecnolog\'ias Afines (CATA) BASAL PFB/06.
R.S. was financed through a combination of GEMINI-CONICYT fund 32080008 and a COMITE MIXTO grant. C.M.B. also
thanks F. Mauro for useful discussions. All authors acknowledge partial support from the Yale
University/Universidad de Chile collaboration. The SPM3 catalog was funded in part by grants from the US
National Science Foundation, Yale University and the Universidad Nacional de San Juan, Argentina. We warmly
thank W.~F. van Altena, V.~I. Korchagin, T.~M. Girard, and D.~I. Casetti-Dinescu for their help and suggestions.

{\it Facilities:} \facility{Du Pont (ECHELLE)}, \facility{Magellan:Clay (MIKE)},
\facility{Euler1.2m (CORALIE)}, \facility{Max Plank:2.2m (FEROS)}


\end{document}